\documentclass[pra,groupedaddress,superscriptaddress,reprint,onecolumn,showpacs,notitlepage,nofootinbib,longbibliography]{revtex4-1}

\usepackage{amsfonts}
\usepackage{amsmath}
\usepackage{amssymb}
\usepackage{mathtools}
\usepackage{graphicx}
\usepackage{enumerate}
\usepackage{wrapfig}
\usepackage{color}
\usepackage{mathrsfs}
\usepackage{bm}
\usepackage{commath}
\usepackage[hidelinks]{hyperref}
\hypersetup{
     colorlinks   = true,
     citecolor    = blue,
     linkcolor    = blue,
     urlcolor     = blue
}
\usepackage{subfigure}
\usepackage{mathptmx}
\usepackage{times}
\usepackage{ulem}
\usepackage{commath}
\usepackage{floatrow}

\DeclareMathAlphabet{\pazocal}{OMS}{zplm}{m}{n}

\DeclareMathOperator{\tr}{tr}

\makeatletter
\DeclareFontFamily{OMX}{MnSymbolE}{}
\DeclareSymbolFont{MnLargeSymbols}{OMX}{MnSymbolE}{m}{n}
\SetSymbolFont{MnLargeSymbols}{bold}{OMX}{MnSymbolE}{b}{n}
\DeclareFontShape{OMX}{MnSymbolE}{m}{n}{
    <-6>  MnSymbolE5
   <6-7>  MnSymbolE6
   <7-8>  MnSymbolE7
   <8-9>  MnSymbolE8
   <9-10> MnSymbolE9
  <10-12> MnSymbolE10
  <12->   MnSymbolE12
}{}
\DeclareFontShape{OMX}{MnSymbolE}{b}{n}{
    <-6>  MnSymbolE-Bold5
   <6-7>  MnSymbolE-Bold6
   <7-8>  MnSymbolE-Bold7
   <8-9>  MnSymbolE-Bold8
   <9-10> MnSymbolE-Bold9
  <10-12> MnSymbolE-Bold10
  <12->   MnSymbolE-Bold12
}{}

\newcommand{\ignore}[1]{}
\newcommand{\nobibentry}[1]{{\let\nocite\ignore\bibentry{#1}}}

\newcommand{\bibfnamefont}[1]{#1}
\newcommand{\bibnamefont}[1]{#1}

\newcommand{\bea}{\begin{eqnarray}}
\newcommand{\eea}{\end{eqnarray}}

\begin{document}

\normalem

\title{Testing the validity of the `local' and `global' GKLS master equations on an exactly solvable model}
\author{J. Onam Gonz\'{a}lez}
\email{jgonzall@ull.es}
\affiliation{Dpto. de F\'{i}sica and IUdEA: Instituto Universitario de Estudios Avanzados, Universidad de La Laguna, 38203 Spain}
\affiliation{School of Mathematical Sciences and Centre for the Mathematics and Theoretical Physics of Quantum Non-Equilibrium Systems, The University of Nottingham, University Park, Nottingham NG7 2RD, United Kingdom}

\author{Luis A. Correa}
\email{luis.correa@nottingham.ac.uk}
\affiliation{School of Mathematical Sciences and Centre for the Mathematics and Theoretical Physics of Quantum Non-Equilibrium Systems, The University of Nottingham, University Park, Nottingham NG7 2RD, United Kingdom}

\author{Giorgio Nocerino}
\email{pmxgn@nottingham.ac.uk}
\affiliation{School of Mathematical Sciences and Centre for the Mathematics and Theoretical Physics of Quantum Non-Equilibrium Systems, The University of Nottingham, University Park, Nottingham NG7 2RD, United Kingdom}

\author{Jos\'{e} P. Palao}
\email{jppalao@ull.es}
\affiliation{Dpto. de F\'{i}sica and IUdEA: Instituto Universitario de Estudios Avanzados, Universidad de La Laguna, 38203 Spain}

\author{Daniel Alonso}
\email{dalonso@ull.es}
\affiliation{Dpto. de F\'{i}sica and IUdEA: Instituto Universitario de Estudios Avanzados, Universidad de La Laguna, 38203 Spain}

\author{Gerardo Adesso}
\email{gerardo.adesso@nottingham.ac.uk}
\affiliation{School of Mathematical Sciences and Centre for the Mathematics and Theoretical Physics of Quantum Non-Equilibrium Systems, The University of Nottingham, University Park, Nottingham NG7 2RD, United Kingdom}

\begin{abstract}

When deriving a master equation for a multipartite weakly-interacting open quantum systems, dissipation is often addressed \textit{locally} on each component, i.e. ignoring the coherent couplings, which are later added `by hand'. Although simple, the resulting local master equation (LME) is known to be thermodynamically inconsistent. Otherwise, one may always obtain a consistent \textit{global} master equation (GME) by working on the energy basis of the full interacting Hamiltonian. Here, we consider a two-node `quantum wire' connected to two heat baths. The stationary solution of the LME and GME are obtained and benchmarked against the exact result. Importantly, in our model, the validity of the GME is constrained by the underlying secular approximation. Whenever this breaks down (for resonant weakly-coupled nodes), we observe that the LME, in spite of being thermodynamically flawed: (a) predicts the correct steady state, (b) yields the exact asymptotic heat currents, and (c) reliably reflects the correlations between the nodes. In contrast, the GME fails at all three tasks. Nonetheless, as the inter-node coupling grows, the LME breaks down whilst the GME becomes correct. Hence, the global and local approach may be viewed as \textit{complementary} tools, best suited to different parameter regimes.

\end{abstract}

\pacs{05.70.-a, 05.30.-d, 03.65.Yz}
\date{\today}
\maketitle

\section{Introduction}\label{sec:introduction}

The Gorini-Kossakowski-Lindblad-Sudarshan (GKLS) quantum master equation \cite{lindblad1976generators,gorini1976completely} is central in the theory of open quantum systems. It reads
\begin{equation}
\frac{\dif\pmb{\varrho}}{\dif t} = \pazocal{L}\pmb{\varrho} = -\frac{i}{\hbar}[\pmb{H},\pmb{\varrho}] + \pazocal{D}\pmb{\varrho} = -\frac{i}{\hbar}[\pmb{H},\pmb{\varrho}] + \sum_k \gamma_k\left( \pmb{A}_k\pmb{\varrho}\pmb{A}_k^\dagger - \frac12\pmb{A}_k^\dagger\pmb{A}_k\pmb{\varrho}-\frac12\pmb{\varrho}\pmb{A}_k^\dagger\pmb{A}_k \right),
\label{eq:lindblad_abstract}
\end{equation}
and generates a quantum \textit{dynamical semigroup}, i.e. it gives rise to a dynamical map $ \pmb{\varrho}(t) = \pazocal{V}(t)\pmb{\varrho}(t_0) = e^{{\pazocal{L}}(t-t_0)}\pmb{\varrho}(t_0) $ with the semi-group property $ \pazocal{V}(t)\pazocal{V}(s) = \pazocal{V}(t+s) $. This type of memoryless or Markovian evolution arises naturally when an open quantum system couples weakly to an environment at inverse-temperature $ \beta = (k_B T)^{-1} $, so that the typical relaxation time is by far the largest scale in the problem \cite{breuer2002theory}.

Among many others, Eq.~\eqref{eq:lindblad_abstract} has the following key properties:
\begin{enumerate}[(i)]
\item It ensures a \textit{completely positive} dynamics which, in turn, implies that the relative entropy $ S(\pmb{\varrho}_1\vert\pmb{\varrho}_2) \coloneqq \tr{\{ \pmb{\varrho}_1(\log{\pmb{\varrho}}_1-\log{\pmb{\varrho}_2}) \}} $ between any two states evolving under $ \pazocal{V}(t) $ decreases \textit{monotonically}, i.e. $ \frac{\dif}{\dif t} S(\pmb{\varrho}_1(t)\vert\pmb{\varrho}_2(t)) \leq 0 $ \cite{spohn1978entropy,muller2015monotonicity}.
\item Under mild assumptions, the thermal state $ \pmb{\tau} \propto \exp{(-\beta\pmb{H})} $ is the \textit{only stationary state} of $ \pazocal{V}(t) $, i.e. $ \pazocal{L}\pmb{\tau} = 0 $ \cite{spohn1977algebraic}. That is, Eq.~\eqref{eq:lindblad_abstract} describes relaxation towards thermal equilibrium.
\end{enumerate}

Interestingly, one may use Eq.~\eqref{eq:lindblad_abstract} to model a continuous (quantum) thermodynamic cycle \cite{alicki1979engine,kosloff1984quantum}. By coupling the open system (i.e. the \textit{working substance}) to various heat baths at different temperatures and possibly also to a periodic external drive, a stationary non-equilibrium state builds up. The direction of the corresponding steady-state heat currents may be controlled by suitably engineering the spectrum of the working substance. Hence, we can speak of `quantum heat engines' or `quantum compression/absorption refrigerators' \cite{PhysRevE.64.056130}, which have attracted a lot of attention in recent years \cite{e15062100,1310.0683v1,gelbwaser2015thermodynamics,vinjanampathy2016quantum,goold2016role}.

In such quantum heat devices, the stationary incoming heat currents $ \{\dot{\pazocal{Q}}_\alpha\} $ and the power output $ -\pazocal{P} $ are defined as \cite{alicki1979engine}
\begin{equation}
\frac{\dif}{\dif t}\tr\left\lbrace \pmb{H}\,\pmb{\varrho}_\infty \right\rbrace = 0 = \pazocal{P} + \sum_\alpha\dot{\pazocal{Q}}_\alpha \coloneqq \tr{\left\lbrace \frac{\partial\pmb{H}}{\partial t}\,\pmb{\varrho}_\infty \right\rbrace} + \sum_{\alpha}\tr{\left\lbrace\pmb{H}\pazocal{D}_\alpha\pmb{\varrho}_\infty\right\rbrace},
\label{eq:heat_work}
\end{equation}
where $ \pmb{\varrho}_\infty $ is the steady state of the working substance, and $ \pazocal{D}_\alpha $ denotes the GKLS dissipation super-operator associated with bath $ \alpha $.

Owing to properties (i) and (ii) above, the stationary heat currents $ \dot{\pazocal{Q}}_\alpha $ satisfy the relation $ \sum_{\alpha}\dot{\pazocal{Q}}_\alpha/T_\alpha \leq 0 $, which is the Clausius inequality. In other words, addressing the dynamics of quantum heat devices with GKLS quantum master equations ensures thermodynamic consistency.

When modelling open quantum systems made up of multiple weakly-interacting parts coupled to local environments, it is commonplace to build the corresponding master equation by simply adding the local dissipators for the relaxation of each individual component (ignoring their coherent interactions). That is, for a multipartite system with Hamiltonian $ \pmb{H} = \sum_j\pmb{h}_j + k\,\pmb{V} $, where $ \pmb{V} $ contains all the internal couplings (of strength $ k $), one would write\footnote{For simplicity, we are omitting the Lamb shift (cf. Sec.~\ref{subsec:global}).}
\begin{equation}
\frac{\dif\pmb{\varrho}}{\dif t} = -\frac{i}{\hbar}[\pmb{H},\pmb{\varrho}] + \sum_{\alpha}\pazocal{D}^{(k=0)}_\alpha\pmb{\varrho}.
\label{eq:local_intro}
\end{equation}

Although Eq.~\eqref{eq:local_intro} is in GKLS form, property (ii) ceases to hold, as the dissipators $ \pazocal{D}^{(k=0)}_\alpha $ do not match the Hamiltonian $ \pmb{H} $, but rather the non-interacting $ \sum_j\pmb{h}_j $. Consequently, describing heat transport with the local master equation \eqref{eq:local_intro} may lead to thermodynamic inconsistencies: Heat could, for instance, flow against the temperature gradient \cite{levy2014local}, or non-vanishing steady-state heat currents could be present even if all reservoirs are set to the same temperature \cite{stockburger2016thermodynamic}.

These observations, strongly advise to follow the standard procedure to consistently obtain the correct global dissipators $ \pazocal{D}_\alpha $ \cite{breuer2002theory}. However, doing so may become particularly challenging when dealing with large systems, e.g. long harmonic or spin chains. Moreover, in such cases the capital assumption that the dissipation time scale is by far the largest in the problem is likely to break down as the spectrum of the system becomes denser \cite{PhysRevE.76.031115}; Eq.~\eqref{eq:lindblad_abstract} would then lack a microscopic justification. These difficulties explain the popularity of simple approaches based on weak internal coupling approximations such as Eq.~\eqref{eq:local_intro} \cite{PhysRevE.76.031115,trushechkin2016perturbative}. In this paper we wish to put such local approaches to the test.

In particular, we choose an exactly solvable model consisting of a two-node harmonic chain weakly coupled on both edges to two heat baths at different temperatures. Our system is set up so that, when the inter-node coupling strength becomes comparable or smaller than the node-baths dissipative couplings, the secular approximation underlying Eq.~\eqref{eq:lindblad_abstract} may break down. This allows us to gauge to which extent the local master equation (LME) remains an accurate description. Interestingly, we find that the local approach yields an excellent approximation to the steady state, the stationary heat currents, and the asymptotic quantum and classical correlations, in the regime of parameters in which the global master equation (GME) fails even qualitatively. More generally, it follows that heat conduction through arbitrarily large harmonic chains can be correctly modelled within the local approach always provided that the internal couplings are sufficiently weak. The present work thus adds to the efforts of Refs.~\cite{rivas2010markovian,1751-8121-43-40-405304,PhysRevE.76.031115,PhysRevLett.111.180602,PhysRevE.94.062143,stockburger2016thermodynamic,levy2014local,trushechkin2016perturbative,PhysRevA.93.062114,deccordi2017two} to clarify the \textit{dos and don'ts} of modelling heat transport through multipartite open quantum systems.

This paper is structured as follows: In Sec.~\ref{subsec:hamiltonian} we outline the steps of the microscopic derivation of the GKLS quantum master equation. We then proceed to derive and solve such an equation for our specific model in Sec~\ref{subsec:global}. The alternative local master equation is obtained in Sec.~\ref{subsec:local}. Before proceeding to benchmark both approaches, in Sec.~\ref{sec:exact} we sketch how the exact steady-state solution of the system may be obtained by solving the quantum Langevin equations. We then devote Sec.~\ref{sec:discussion} to present and discuss our results. Finally, in Sec.~\ref{sec:conclusions} we summarize and draw our conclusions.

\section{Deriving Markovian master equations}\label{sec:master_eqs}

\subsection{The model, the Markovian master equation and its steady state}\label{subsec:hamiltonian}

\begin{figure}
	\floatbox[{\capbeside\thisfloatsetup{capbesideposition={right,top},capbesidewidth=0.45\textwidth}}]{figure}[\FBwidth]
	{\caption{Schematic representation of the wire. The two harmonic nodes at frequencies $ \omega_c $ and $ \omega_h $ are coupled through a spring-like interaction of strength $ k $. Each node is, in turn, dissipatively coupled to a `cold' and `hot' heat bath at temperatures $ T_c < T_h $. The dissipation strength $ \lambda^2 $ is assumed sufficiently weak to justify the use of a perturbative master equation up to $ \pazocal{O}(\lambda^2) $.}\label{fig0}}
	{\includegraphics[width=0.45\textwidth]{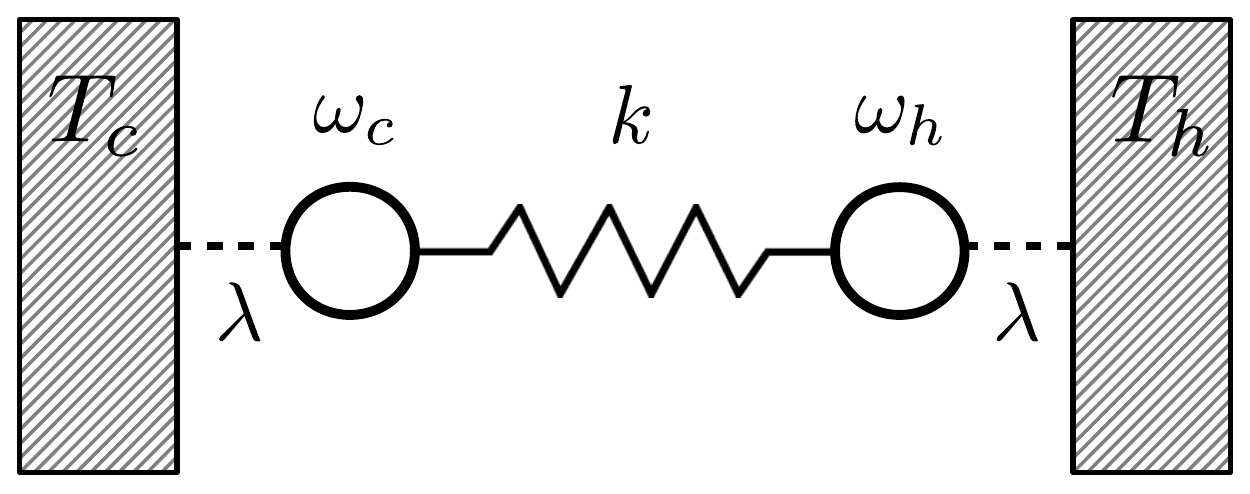}}
\end{figure}

We will consider a two-node `quantum wire' (see Fig.~\ref{fig0}) consisting of mechanically-coupled harmonic oscillators with bare frequencies $ \omega_c $ and $ \omega_h $ and coupling strength $ k>0 $. Each node will be weakly connected to a bosonic bath, i.e. an infinite collection of uncoupled harmonic modes in thermal equilibrium (at temperatures $ T_c < T_h $). The total Hamiltonian may be cast as
\begin{equation}
\pmb{H} = \sum_{\alpha\in\{c,h\}}\left(\frac{\omega_\alpha^2}{2}\pmb{X}_\alpha^2 + \frac{\pmb{P}_\alpha^2}{2}\right) + \frac{k}{2}\left(\pmb{X}_c-\pmb{X}_h\right)^2 + \sum_{\alpha\in\{c,h\}}\sum_{\mu}\left(\frac{\omega_{\alpha,\mu}^2 m_{\alpha,\mu}}{2}\pmb{x}_{\alpha,\mu}^2+\frac{\pmb{p}_{\alpha,\mu}^2}{2 m_{\alpha,\mu}}\right) - \sum_{\alpha\in\{c,h\}}\pmb{X}_\alpha\otimes\sum_{\mu} g_{\alpha,\mu} \pmb{x}_{\alpha,\mu},
\label{eq:hamiltonian}
\end{equation}
where the masses of the nodes have been set to $ m_c = m_h = 1 $, and the constants $ g_{\alpha,\mu} $ stand for the coupling strength between node $ \alpha $ and each of the environmental modes $(\alpha,\mu)$. Also, in all what follows we shall set $ \hbar $ and the Boltzmann constant $ k_B $ to $ 1 $. We will refer to the first three terms in the right-hand side of Eq.~\eqref{eq:hamiltonian} as the free (system + baths) Hamiltonian $ \pmb{H}_0 = \pmb{H}_S + \pmb{H}_B $, as opposed to the last term $ \pmb{H}_\text{int} $, which describes the system-baths interaction. For later convenience, we shall also introduce the notation $ \pmb{B}_\alpha\coloneqq\sum_{\mu}g_{\alpha,\mu} \pmb{x}_{\alpha,\mu}$.

We will group the system-baths cupling constants in the \textit{spectral density} functions defined as $ J_\alpha(\omega) \coloneqq \pi\,\sum_\mu \frac{g_{\alpha,\mu}^2}{2 m_\mu \omega_\mu} \delta(\omega-\omega_\mu) $. In particular, we will choose 1D baths with the Ohmic form
\begin{equation}
J_{c}(\omega) = J_{h}(\omega) = \lambda^2\,\omega \frac{\Lambda^2}{\omega^2+\Lambda^2},
\label{eq:spectraldensity}
\end{equation}
where $ \Lambda $ is a high-frequency cutoff $ (\max\{\omega_c,\omega_h\} \ll \Lambda) $ and the parameter $ \lambda $ captures the \textit{dissipation strength}. Note that the bath operators $ \pmb{B}_\alpha $ are thus $ \pazocal{O}(\lambda) $.

For completeness, we will now briefly sketch a simple procedure to obtain the standard second-order Markovian generator for the reduced dynamics of the system (see Ref.~\cite{breuer2002theory} for full details). Let us take the Liouville-von Neumann equation in the interaction picture
\begin{equation}
\frac{\dif\tilde{\pmb{\rho}}(t)}{\dif t} = -i[\tilde{\pmb{H}}_\text{int}(t),\tilde{\pmb{\rho}}(t)],
\label{eq:Liouville-vonNeumann}
\end{equation}
where $ \tilde{\pmb{H}}_\text{int}(t)\coloneqq e^{i\,\pmb{H}_0 t}\,\pmb{H}_\text{int} e^{-i\,\pmb{H}_0 t}$t and $ [\cdot,\cdot] $ stands for a commutator. Formally integrating Eq.~\eqref{eq:Liouville-vonNeumann} and assuming that the initial condition is such that $ \tr\{\tilde{\pmb{\rho}}(0)\,\tilde{\pmb{H}}_\text{int}\} = 0 $ yields the following equation of motion for the system:
\begin{equation}
\frac{\dif\tilde{\pmb{\sigma}}}{\dif t} = -\int_0^t\dif s\,\tr_B[\tilde{\pmb{H}}_\text{int}(t),[\tilde{\pmb{H}}_\text{int}(s),\tilde{\pmb{\rho}}(s)]].
\label{eq:Liouville-vonNeumann-integrated}
\end{equation}

Here, $ \tilde{\pmb{\sigma}}\coloneqq\tr_B\tilde{\pmb{\rho}} $ and $ \tr_B\{\cdots\} $ denotes trace over the baths. We will now assume that the dissipation strength $ \lambda $ is so weak that when starting from a factorized initial condition $ \pmb{\rho}_0\coloneqq\pmb{\sigma}(0)\otimes\pmb{\tau}_c\otimes\pmb{\tau}_h $ the propagated state $ \tilde{\pmb{\rho}}(t)\simeq\tilde{\pmb{\sigma}}(t)\otimes\pmb{\tau}_c\otimes\pmb{\tau}_h $ remains approximately factorized at all times. $ \pmb{\tau}_{\alpha\in\{c,h\}} $ are thermal states for the hot and cold bath. We will also replace $ \tilde{\pmb{\sigma}}(s) $ inside the integral in Eq.~\eqref{eq:Liouville-vonNeumann-integrated} by $ \tilde{\pmb\sigma}(t) $, thus making it time-local. The change of variables $ s\rightarrow t-s $ yields the Redfield equation \cite{redfield1957theory,RevModPhys.89.015001}
\begin{equation}
\frac{\dif\tilde{\pmb{\sigma}}}{\dif t} \simeq -\int_0^t\dif s\,\tr_B[\tilde{\pmb{H}}_\text{int}(t),[\tilde{\pmb{H}}_\text{int}(t-s),\tilde{\pmb{\sigma}}(t)\otimes\pmb{\tau}_c\otimes\pmb{\tau}_h]].
\label{eq:RedfieldRaw}
\end{equation}
Notice that the resulting state $ \pmb{\varrho}(t) $ does keep a memory of the initial condition $ \pmb{\varrho}(0) $ and hence, Eq.~\eqref{eq:RedfieldRaw} is non-Markovian. However, provided that the integrand above decays sufficiently fast, one might set $ t\rightarrow\infty $ in the upper limit of integration, which is referred-to as Born-Markov approximation. This approximation is justified whenever $ \lambda^2 \ll \min\{ T, \Lambda \} $.

A further step still remains to be undertaken in order to bring the resulting `Markovian Redfield' master equation to the canonical GKLS form---the secular approximation. Let us examine $ \tilde{\pmb{H}}_\text{int}(t) $ more closely. One may always decompose $ \pmb{X}_\alpha = \sum_{\omega}\pmb{L}_\alpha^\omega $, where $ [\pmb{H}_S,\pmb{L}_\alpha^\omega] = -\omega\,\pmb{L}_\alpha^\omega$, so that $ \tilde{\pmb{H}}_\text{int} = \sum_\omega e^{-i\omega \,t}\pmb{L}_\alpha^\omega\otimes\tilde{\pmb{B}}_\alpha(t) $, with the interaction-picture bath operator $ \tilde{\pmb{B}}_\alpha(t) = e^{i\pmb{H}_B t}\pmb{B}_\alpha e^{-i\pmb{H}_B t} $. Plugging this into Eq.~\eqref{eq:RedfieldRaw} leads to
\begin{equation}
\frac{\dif \tilde{\pmb{\sigma}}}{\dif t} \simeq\frac12 \sum_\alpha\sum_{\omega,\omega'}e^{i(\omega'-\omega)t}\gamma_\alpha(\omega)\left(\pmb{L}_\alpha^\omega\,\tilde{\pmb{\sigma}}\,{\pmb{L}^{\omega'}_\alpha}^\dagger-{\pmb{L}^{\omega'}_\alpha}^\dagger\,\pmb{L}_\alpha^\omega\,\tilde{\pmb{\sigma}}\right) + \text{h.c.},
\label{eq:RedfieldCooked}
\end{equation}
where $ \gamma_\alpha(\omega) = 2\text{Re}\,\int_0^\infty\dif s\, e^{i\omega\,s}\tr_B\{\pmb{B}_\alpha(t)\,\pmb{B}_\alpha(t-s)\} $. Note that we are completely ignoring $ \text{Im}\,\int_0^\infty\dif s\, e^{i\omega\,s}\tr_B\{\pmb{B}_\alpha(t)\,\pmb{B}_\alpha(t-s)\} $, which would eventually lead to a mere shift (of order $ \lambda^2 $) on the energy levels of the Hamiltonian (Lamb shift) \cite{breuer2002theory}. The secular approximation consists in time-averaging Eq.~\eqref{eq:RedfieldCooked} over an interval of the order of the dissipation time $ T_D \sim \lambda^{-2} $. All terms with $ \omega'\neq\omega $ above can then be discarded provided that they oscillate fast as compared with $ T_D $. Returning to the Schr\"{o}dinger picture, finally leaves us with the GKLS quantum master equation
\begin{equation}
\frac{\dif \pmb{\sigma}}{\dif t} \simeq -i[\pmb{H}_S,\pmb{\sigma}] + \sum_\alpha\pazocal{D}_\alpha\pmb{\sigma} = -i[\pmb{H}_S,\pmb{\sigma}] + \sum_\alpha\sum_\omega \gamma_\alpha(\omega)\left(\pmb{L}_\alpha^\omega\,\pmb{\sigma}\,{\pmb{L}_\alpha^\omega}^\dagger-\frac12\{{\pmb{L}_\alpha^\omega}^\dagger\,\pmb{L}_\alpha^\omega,\pmb{\sigma}\}_+\right),
\label{eq:GKLS}
\end{equation}
where $ \{\cdot,\cdot\}_+ $ stands for anti-commutator. In the next section, we will concentrate in obtaining the specific form of the operators $ \pmb{L}_\alpha^\omega $ for the Hamiltonian in Eq.~\eqref{eq:hamiltonian}.

Because Eq.~\eqref{eq:hamiltonian} is overall quadratic in position and momenta, the steady state will be Gaussian and thus, fully characterized by its first and second order moments \cite{GaussianOSID}. In fact, one can easily see that $ \langle\pmb{X}_\alpha\rangle = \langle\pmb{P}_\alpha\rangle = 0 $, where $ \langle\cdot\rangle $ denotes stationary average. As a result, the steady state will be specified only by the $ 4\times4 $ \textit{covariance matrix}, with elements $ [\mathsf{\Gamma}]_{kl} \coloneqq \frac12\langle \{\pmb{R}_k,\pmb{R}_l\}_+\rangle $, with $ \vec{\pmb{R}} = (\pmb{X}_c, \pmb{P}_c, \pmb{X}_h, \pmb{P}_h )^\mathsf{T} $.

Since we wish to calculate the covariances $ [\mathsf{\Gamma}]_{kl} $ rather than the state $ \pmb{\sigma} $, it will be more convenient to work with the \textit{adjoint} master equation \cite{breuer2002theory} which, for an arbitrary system observable in the Heisenberg picture $ \pmb{O}(t) $, reads
\begin{equation}
\frac{\dif\pmb{O}}{\dif t} \simeq i[\pmb{H}_S,\pmb{O}] + \sum_\alpha\pazocal{D}_\alpha^\dagger\pmb{O} = i[\pmb{H}_S,\pmb{O}] + \sum_\alpha\sum_\omega \gamma_\alpha(\omega)\left({\pmb{L}_\alpha^\omega}^\dagger\,\pmb{O}\,\pmb{L}_\alpha^\omega-\frac12\{{\pmb{L}_\alpha^\omega}^\dagger\,\pmb{L}_\alpha^\omega,\pmb{O}\}_+\right).
\label{eq:adjoint_GKLS}
\end{equation}

\subsection{The global master equation}\label{subsec:global}

The first step to derive consistent $ \pmb{L}_\alpha^\omega $ operators will be to rotate $ \pmb{H}_S $ into its normal modes. These are
\begin{subequations}
\begin{align}
\pmb{\eta}_+ &= \cos\vartheta\,\pmb{X}_c-\sin\vartheta\,\pmb{X}_h \\
\pmb{\eta}_- &= \sin\vartheta\,\pmb{X}_c+\cos\vartheta\,\pmb{X}_h,
\end{align}
\label{eq:coordinates+-}
\end{subequations}
where the angle $ \vartheta $ is
\begin{equation}
\cos^2\vartheta = \frac{-\delta_\omega^2+\sqrt{4k^2+\delta_\omega^4}}{2\sqrt{4k^2+\delta_\omega^4}}
\label{eq:theta}
\end{equation}
and, in turn, $ \delta_\omega^2\coloneqq\omega_h^2-\omega_c^2 $. The corresponding normal-mode frequencies write as
\begin{equation}
\varOmega^2_\pm = \frac12 \left( \omega_c^2 + \omega_h^2 + 2k\pm\sqrt{4k^2+\delta_\omega^4} \right).
\label{eq:normal_mode_freq}
\end{equation}

After this transformation, the Hamiltonian \eqref{eq:hamiltonian} rewrites as
\begin{equation}
\pmb{H} = \sum_{s\in\{+,-\}}\left(\frac{\varOmega_s^2}{2}\pmb{\eta}_s^2+\frac{\pmb{\Pi}_s^2}{2}\right) + \pmb{H}_B - (\cos\vartheta\,\pmb{\eta}_+ + \sin\vartheta\,\pmb{\eta}_-)\otimes \pmb{B}_c + (\sin\vartheta\,\pmb{\eta}_+ - \cos\vartheta\,\pmb{\eta}_-)\otimes\pmb{B}_h,
\label{eq:normal_mode_hamiltonian}
\end{equation}
where $ \pmb{\Pi}_s = \dif\pmb{\eta}_s/\dif t$. By writing $ \pmb{\eta}_s = (\pmb{a}_s + \pmb{a}_s^\dagger)/\sqrt{2\varOmega_s} $ (with $ \pmb{a}_s $ and $ \pmb{a}^\dagger_s $ being annihilation and creation operators on mode $ \varOmega_s $) one can see that $ \pmb{X}_\alpha = \pmb{L}_\alpha^{\varOmega_+} + \pmb{L}_\alpha^{\varOmega_-} + \text{h.c.} $, where $ \pmb{L}_c^{\varOmega_+} \coloneqq \cos{\vartheta}\,\pmb{a}_+/\sqrt{2\varOmega_+} $, $ \pmb{L}_c^{\varOmega_-} \coloneqq \,\sin\vartheta\,\pmb{a}_-/\sqrt{2\varOmega_-} $, $\pmb{L}_h^{\varOmega_+}\coloneqq-\,\sin\vartheta\,\pmb{a}_+/\sqrt{2\varOmega_+} $, $ \pmb{L}_h^{\varOmega_-} \coloneqq \,\cos\vartheta\,\pmb{a}_-/\sqrt{2\varOmega_-} $, and $ \pmb{L}_\alpha^{-\varOmega_\pm}\coloneqq{\pmb{L}_\alpha^{\varOmega_\pm}}^\dagger $.

Looking back to the right-hand side of Eq.~\eqref{eq:RedfieldCooked}, we see that there are 16 terms associated with $ 5 $ different open decay channels, oscillating as $e^{i(\omega'-\omega)t}  $ at frequencies $ \vert\omega'-\omega\vert = \{0,2\varOmega_+,2\varOmega_-,\varOmega_+ + \varOmega_-,\varOmega_+-\varOmega_-\} $. Provided that the nodes are sufficiently detuned, i.e. $ \delta_\omega \gg \lambda^2 $, the secular approximation is guaranteed to be valid for \textit{any} value of the coupling $ k $. However, if $ \omega_h-\omega_c $ became comparable or smaller than the dissipation strength $ \lambda^2 $, there would be no justification to discard the non-secular terms oscillating at $ \varOmega_+ - \varOmega_- $ when $ k $ becomes very small. Indeed, defining $ R_\pm \coloneqq 2k\pm\sqrt{4k^2 + \delta^4_\omega} $ one may write
\begin{equation}
\varOmega_+-\varOmega = \sqrt{\frac{\omega_c^2+\omega_h^2}{2}}\left( \sqrt{1+\frac{R_+}{\omega_c^2 + \omega_h^2}} - \sqrt{1+\frac{R_-}{\omega_c^2 + \omega_h^2}} \right)
\label{eq:freq_diff}
\end{equation}
whenever $ R_\pm/(\omega_c^2 + \omega_h^2) \ll 1 $, the Taylor expansion $ \sqrt{1 + x} = 1 +\frac{x}{2} - \frac{x^2}{8} + \cdots $ allows to approximate Eq.~\eqref{eq:freq_diff} as
\begin{equation}
\varOmega_+-\varOmega_- = \sqrt{\frac{4k^2 + \delta_\omega^4}{2(\omega_h^2 + \omega_c^2)}} + \pazocal{O}\left( \frac{R_+^2}{(\omega_c^2 + \omega_h^2)^{3/2}}\right).
\label{eq:fre_diff_approx}
\end{equation}
From Eq.~\eqref{eq:fre_diff_approx} it is clear that for the secular approximation to be valid the dissipation rate must be such that
\begin{equation}
 \lambda^2 \ll \sqrt{\frac{4k^2 + \delta_\omega^4}{2(\omega_h^2 + \omega_c^2)}},
\label{eq:validity_secular}
\end{equation}
which, in the limit of resonant nodes simplifies to $ \lambda^2 \ll k/\omega_c $. Hence, we can anticipate that Eq.~\eqref{eq:GKLS} will fail to describe nearly resonant weakly coupled nodes, which is precisely the regime in which we shall focus our analysis.

The only additional ingredient required to build Eq.~\eqref{eq:adjoint_GKLS} are the \textit{decay rates} $ \gamma_\alpha(\pm\varOmega_\pm) $. A direct calculation leads to
\begin{equation}
\gamma_\alpha(\omega) = 2 J(\omega)[1+n_\alpha(\omega)],
\label{eq:decay_rates}
\end{equation}
where $ n_\alpha(\omega) \coloneqq (e^{\omega/T_\alpha}-1)^{-1}$ is the bosonic occupation number for frequency $ \omega $ at temperature $ T_\alpha $. Note that $ \gamma_\alpha(-\omega) = \exp{(-\omega/T_\alpha)} \gamma_\alpha(\omega) $. Combining all the above, and after tedious but straightforward algebra, we can obtain a closed set of equations of motion for the covariances\footnote{It is indeed enough to consider the equations of motion for the mode occupation numbers $ \langle\pmb{a}_\pm^\dagger\pmb{a}_\pm\rangle $ (which are decoupled), to fully solve the dynamics.} $ \langle\pmb{\eta}_\pm^2\rangle $, $ \langle\pmb{\Pi}_\pm^2\rangle $, and $ \langle\{\pmb{\eta}_\pm,\pmb{\Pi}_\pm\}_+\rangle $. Note that $ \langle\cdot\rangle $ denotes here instantaneous average.
\begin{subequations}
\begin{align}
&\frac{\dif}{\dif t} \langle \pmb{\eta}_{\pm}^2  \rangle = \varDelta_{\pm} \langle \pmb{\eta}_{\pm}^2  \rangle + \langle \{\pmb{\eta}_\pm,\pmb{\Pi}_\pm \}_+\rangle + \frac{\varSigma_{\pm}}{2\varOmega_{\pm}}                  \\
&\frac{\dif}{\dif t} \langle \{\pmb{\eta}_\pm,\pmb{\Pi}_\pm\}_+ \rangle = 2\langle\pmb{\Pi}_\pm^2\rangle - 2\varOmega_\pm\langle\pmb{\eta}_\pm^2\rangle + \varDelta_{\pm}\langle \{\pmb{\eta}_\pm,\pmb{\Pi}_\pm\}_+ \rangle\\
&\frac{\dif}{\dif t} \langle \pmb{\Pi}_{\pm}^2  \rangle = \varDelta_{\pm} \langle \pmb{\Pi}_{\pm}^2  \rangle -\varOmega_\pm^2\langle\{\pmb{\eta}_\pm,\pmb{\Pi}_\pm\}_+\rangle+\frac{\varOmega_{\pm}}{2}\varSigma_{\pm},
\end{align}
\label{eq:global}
\end{subequations}
where the following notations have been introduced
\begin{subequations}
\begin{align}
\varDelta_{+} & \coloneqq \frac{\cos^2\vartheta}{2\varOmega_+}\,[\gamma_c(-\varOmega_+) - \gamma_c(\varOmega_+)]  +  \frac{\sin^{2}\vartheta}{2\varOmega_+}\,[\gamma_h(-\varOmega_+) - \gamma_h(\varOmega_+)] \\
\varDelta_{-} & \coloneqq \frac{\sin^2\vartheta}{2\varOmega_-}\,[\gamma_c(-\varOmega_-) - \gamma_c(\varOmega_-)]  +  \frac{\cos^{2}\vartheta}{2\varOmega_-}\,[\gamma_h(-\varOmega_-) - \gamma_h(\varOmega_-)] \\
\varSigma_{+} & \coloneqq \frac{\cos^2\vartheta}{2\varOmega_+}\,[\gamma_c(-\varOmega_+) + \gamma_c(\varOmega_+)]  +  \frac{\sin^{2}\vartheta}{2\varOmega_+}\,[\gamma_h(-\varOmega_+) + \gamma_h(\varOmega_+)] \\
\varSigma_{-} & \coloneqq \frac{\sin^2\vartheta}{2\varOmega_-}\,[\gamma_c(-\varOmega_-) + \gamma_c(\varOmega_-)]  +  \frac{\cos^{2}\vartheta}{2\varOmega_-}\,[\gamma_h(-\varOmega_-) + \gamma_h(\varOmega_-)].
\end{align}
\label{eq:coefficients}
\end{subequations}
Below, it will be convenient to break down each of these coefficients into the sum of its two constituent terms, as $ \varDelta_{\pm}\coloneqq\varDelta_\pm^c + \varDelta_\pm^h $ and $\varSigma_{\pm} \coloneqq \varSigma_\pm^c + \varSigma_\pm^h$, where e.g. $ \varDelta_+^c =  \cos^2\vartheta\,[\gamma_c(-\varOmega_+) - \gamma_c(\varOmega_+)]/(2\varOmega_+) $. The further denotations $ \varSigma_\pm^\alpha = W^\alpha_{-\varOmega_\pm} + W^\alpha_{\varOmega_\pm} $, where e.g. $ W^c_{-\varOmega_+} = \cos^2\vartheta\gamma_c(-\varOmega_+)/(2\varOmega_+) $, will also be employed later on.

The stationary solution to Eq.~\eqref{eq:global} is simply $ \langle \pmb{\eta}_\pm^2 \rangle = -\varSigma_\pm/(2\varDelta_\pm\varOmega_\pm) $, $ \langle\pmb{\Pi}_\pm\rangle = -\varOmega_\pm\varSigma_\pm/(2\varDelta\pm) $, and $ \langle\{\pmb{\eta}_\pm,\pmb{\Pi}_\pm\}_+\rangle = 0 $, so that the non-zero elements of the asymptotic covariance matrix in the original quadratures reads
\begin{equation}
\mathsf{\Gamma}^\text{G} = \begin{pmatrix}
[\mathsf{\Gamma}^\text{G}]_{11} && 0 && [\mathsf{\Gamma}^\text{G}]_{13} && 0 \\
0 && [\mathsf{\Gamma}^\text{G}]_{22} && 0 && [\mathsf{\Gamma}^\text{G}]_{24} \\
[\mathsf{\Gamma}^\text{G}]_{13} && 0 && [\mathsf{\Gamma}^\text{G}]_{33} && 0 \\
0 && [\mathsf{\Gamma}^\text{G}]_{24} && 0 && [\mathsf{\Gamma}^\text{G}]_{44}
\end{pmatrix},
\label{eq:cov_global}
\end{equation}
where $ [\mathsf{\Gamma}^\text{G}]_{11} = \langle\pmb{\eta}_+^2\rangle\cos^2\vartheta + \langle\pmb{\eta}_-^2\rangle\sin^2\vartheta $, $
[\mathsf{\Gamma}^\text{G}]_{22} = \langle \pmb{\Pi}_+^2 \rangle\cos^2\vartheta + \langle \pmb{\Pi}_-^2 \rangle\sin^2\vartheta $, $
[\mathsf{\Gamma}^\text{G}]_{33} = \langle\pmb{\eta}_+^2\rangle\sin^2\vartheta + \langle\pmb{\eta}_-^2\rangle\cos^2\vartheta $, $
[\mathsf{\Gamma}^\text{G}]_{44} = \langle \pmb{\Pi}_+^2 \rangle\sin^2\vartheta + \langle \pmb{\Pi}_-^2 \rangle\cos^2\vartheta $, $
[\mathsf{\Gamma}^\text{G}]_{13} = (\langle\pmb{\eta}^2_-\rangle-\langle\pmb{\eta}^2_+\rangle)\sin\vartheta\,\cos\vartheta $, and $
[\mathsf{\Gamma}^\text{G}]_{24} = (\langle\pmb{\Pi}_-^2\rangle-\langle\pmb{\Pi}_+^2\rangle)\sin\vartheta\,\cos\vartheta $. Finally, the steady-state heat currents can be written as
\begin{equation}
\dot{\pazocal{Q}}_\alpha^\text{G} = \tr\{ \pmb{H}_S\pazocal{D}_\alpha\pmb{\sigma}(\infty) \} = \langle\pazocal{D}_\alpha^\dagger\,\pmb{H}_S\rangle = \frac12\sum_{s\in\{+,-\}}\left[\varDelta_s^\alpha\left(\varOmega_s^2\langle\pmb{\eta}_s^2\rangle + \langle\pmb{\Pi}_s^2\rangle\right) + \varOmega_s\varSigma_s^\alpha\right].
\label{eq:heat_GME}
\end{equation}
Using Eqs.~\eqref{eq:coefficients} we can cast \eqref{eq:heat_GME} as
\begin{equation}
\dot{\pazocal{Q}}_h^\text{G} = - \dot{\pazocal{Q}}_c^\text{G} = \sum_{s\in\{+,-\}}\varOmega_s\frac{W^c_{\varOmega_s}W^h_{\varOmega_s}}{\varSigma_s}(e^{-\varOmega_s/T_h}-e^{-\varOmega_s/T_c}),	
\label{eq:heat_GME_sign}
\end{equation}
from where it is clear that $ T_h > T_c $ entails $ \dot{\pazocal{Q}}_h^\text{G} = - \dot{\pazocal{Q}}_c^\text{G} > 0 $; that is, heat \textit{always} flows from the hotter bath into the colder one.

\subsection{The local master equation}\label{subsec:local}

Recall from Sec.~\ref{sec:introduction} that, while the \textit{local} master equation looks formally identical to the GME, the choice of operators $ \pmb{L}_\alpha^\omega $ in the local approach is not consistent with the Hamiltonian $ \pmb{H}_S $. As already advanced and provided that the coupling $ k $ is weak, one could derive two independent local dissipators $ \pazocal{D}_\alpha^{(k=0)} $, acting on the cold and hot nodes \textit{separately}, to then construct an approximate equation of motion such as $ \dif \pmb{\sigma}/\dif t \simeq -i[\pmb{H}_S,\pmb{\sigma}] + \sum_{\alpha\in\{c,h\}}\pazocal{D}_\alpha^{(k=0)}\pmb{\sigma} $, as an alternative to Eqs.~\eqref{eq:global}. One might argue that this is a convenient strategy whenever finding all energy eigenstates of the full interacting Hamiltonian is hard, as these are required to write the decomposition $ \{\pmb{L}_\alpha^\omega\} $ of the system operator coupled to each bath \cite{trushechkin2016perturbative}. In our simple example, however, the local approach leads to a more complicated dynamics than the global one---all $ 10 $ independent covariances are needed in order to obtain a closed set of equations of motion.

Specifically, within the local approach one decomposes $ \pmb{X}_\alpha = \pmb{L}_\alpha^{\omega_\alpha} + \pmb{L}_\alpha^{-\omega_\alpha} $, where $ \pmb{L}_\alpha^{\omega_\alpha} \coloneqq \,\pmb{b}_\alpha/\sqrt{2\omega_\alpha} $, $ \pmb{b}_\alpha $ is an annihilation operator on node $ \omega_\alpha $, and $ \pmb{L}_\alpha^{-\omega_\alpha} \coloneqq {\pmb{L}_\alpha^{\omega_\alpha}}^\dagger $. The adjoint master equation Eq.~\eqref{eq:adjoint_GKLS} thus becomes
\begin{equation}
\frac{\dif\pmb{O}}{\dif t} \simeq i[\pmb{H}_S,\pmb{O}] + \sum_{\alpha\in\{c,h\}}\left[\frac{\gamma_\alpha(\omega_\alpha)}{2\omega_\alpha}\left(\pmb{b}_\alpha^\dagger\pmb{O}\,\pmb{b}_\alpha-\frac12\{\pmb{b}_\alpha^\dagger\pmb{b}_\alpha,\pmb{O}\}_+\right) + \frac{\gamma_\alpha(-\omega_\alpha)}{2\omega_\alpha}\left(\pmb{b}_\alpha\,\pmb{O}\,\pmb{b}_\alpha^\dagger-\frac12\{\pmb{b}_\alpha\pmb{b}_\alpha^\dagger,\pmb{O}\}_+\right)\right].
\label{eq:local_formal}
\end{equation}

From \eqref{eq:local_formal}, the equations of motion for the elements of the corresponding covariance matrix $ \mathsf{\Gamma}^\text{L} $ are found to be
\begin{subequations}
\begin{align}
&\frac{\dif}{\dif t}\langle \pmb{X}^2_\alpha \rangle = \langle \{\pmb{X}_\alpha,\pmb{P}_\alpha\}_+ \rangle + \tilde{\varDelta}_\alpha \langle\pmb{X}_\alpha^2 \rangle + \frac{\tilde{\varSigma}_\alpha}{2\omega_\alpha} \\
&\frac{\dif}{\dif t}\langle \pmb{P}^2_\alpha \rangle  = 2k \langle \pmb{X}_{\bar{\alpha}}\pmb{P}_\alpha\rangle- \nu_{\alpha}^2 \langle \{\pmb{X}_\alpha,\pmb{P}_{\alpha}\}_+ \rangle + \tilde{\varDelta}_{\alpha}\langle \pmb{P}^2_\alpha \rangle + \frac{\omega_\alpha \tilde{\varSigma}_\sigma}{2} &(\bar{\alpha}\neq \alpha) \\
&\frac{\dif}{\dif t} \langle\{\pmb{X}_\alpha,\pmb{P}_{\alpha}\}\rangle = 2\langle \pmb{P}^2_\alpha \rangle + \tilde{\varDelta}_\alpha \langle\{\pmb{X}_\alpha\pmb{P}_{\alpha}\}_+ \rangle - 2\nu_\alpha^2 \langle \pmb{X}^2_\alpha \rangle  + 2k \langle \pmb{X}_\alpha\pmb{X}_{\bar{\alpha}} \rangle  &(\bar{\alpha}\neq \alpha) \\
&\frac{\dif}{\dif t} \langle \pmb{X}_\alpha \pmb{P}_{\bar{\alpha}} \rangle  = \langle \pmb{P}_\alpha \pmb{P}_{\bar{\alpha}} \rangle + k \langle \pmb{X}^2_\alpha \rangle +\frac{1}{2}( \tilde{\varDelta}_\alpha + \tilde{\varDelta}_{\bar{\alpha}}) \langle \pmb{X}_\alpha\pmb{P}_{\bar{\alpha}} \rangle - \nu^2_{\bar{\alpha}} \langle \pmb{X}_\alpha\pmb{X}_{\bar{\alpha}} \rangle &(\bar{\alpha}\neq \alpha) \\
&\frac{\dif}{\dif t}\langle \pmb{X}_c\pmb{X}_h \rangle = \langle \pmb{X}_c\pmb{P}_h \rangle + \langle \pmb{X}_h\pmb{P}_c \rangle + \frac{1}{2} ( \tilde{\varDelta}_c + \tilde{\varDelta}_h )\langle \pmb{X}_c\,\pmb{X}_h \rangle  \\
&\frac{\dif}{\dif t} \langle \pmb{P}_c\pmb{P}_h \rangle = \frac{k}{2} \left(\langle \{\pmb{X}_h,\pmb{P}_h\}_+ \rangle + \langle \{\pmb{X}_c,\pmb{P}_c\}_+  \rangle\right) -\nu^2_{c}\langle \pmb{X}_c\pmb{P}_h \rangle - \nu^2_{h}\langle \pmb{X}_h\pmb{P}_c \rangle + \frac{1}{2}( \tilde{\varDelta}_{c} + \tilde{\varDelta}_{h} ) \langle \pmb{P}_c\pmb{P}_h \rangle
\end{align}
\label{eq:local}
\end{subequations}
where $\nu_\alpha^2 \coloneqq \omega_{\alpha}^2 + k$, $\tilde{\varDelta}_{\alpha} \coloneqq \frac{\gamma_\alpha(-\omega_\alpha)-\gamma_\alpha(\omega_\alpha)}{2\omega_{\alpha}}$, and $\tilde{\varSigma}_\alpha \coloneqq \frac{\gamma_\alpha(-\omega_\alpha) + \gamma_\alpha(\omega_\alpha)}{2\omega_{\alpha}} $, and the angled brackets denote again instantaneous average. The stationary solution of Eq.~\eqref{eq:local} is cumbersome but the steady-state heat currents can be compactly cast as
\begin{equation}
\dot{\pazocal{Q}}_\alpha^\text{L} = \frac{\tilde{\varDelta}_\alpha}{2}\left[\omega_\alpha^2\langle\pmb{X}_\alpha^2\rangle + \langle\pmb{P}_\alpha^2\rangle+k(\langle\pmb{X}_\alpha^2\rangle - \langle\pmb{X}_\alpha\pmb{X}_{\tilde{\alpha}}\rangle)\right]+\frac{\tilde{\varSigma}_\alpha}{2}\left(\omega_\alpha + \frac{k}{2\omega_\alpha} \right).
\label{eq:heat_LME}
\end{equation}

As anticipated above and unlike Eq.~\eqref{eq:global}, Eq.~\eqref{eq:local} does not necessarily yield a thermodynamically consistent steady state: One could even encounter striking situations for which $ \dot{\pazocal{Q}}_h^\text{L} = -\dot{\pazocal{Q}}_c^\text{L} < 0 $ for $ T_h > T_c $ or $ \dot{\pazocal{Q}}_\alpha^\text{L}\neq 0 $ for $ T_h=T_c $, as illustrated in \cite{levy2014local,stockburger2016thermodynamic}.

\subsection{Comment on the general validity of the local approach for modelling heat transport under weak internal coupling}\label{subsec:comment}

In spite of its thermodynamic inconsistencies, as it was pointed out in Ref.~\cite{trushechkin2016perturbative} the LME \eqref{eq:local_formal} can be formally understood as the lowest order in the perturbative expansion $ \pazocal{D}_\alpha = \pazocal{D}_\alpha^{(0)} + \pazocal{D}_\alpha^{(1)} k + \pazocal{D}^{(2)}_\alpha k^2 + \cdots $, where $ \pazocal{D}_\alpha^{(0)} = \pazocal{D}_\alpha^{(k=0)} $. The LME \eqref{eq:local_formal} would therefore be correct up to $ \pazocal{O}(\lambda^2 k) $ and any thermodynamic inconsistency encountered should fall within this `error bar'.

Note that the GME is itself a perturbative master equation which neglects corrections of order $ \pazocal{O}(\lambda^3) $ and below \cite{gaspard1999slippage}. However, it is guaranteed to give rise to thermodynamically consistent steady-state heat currents \cite{alicki1979engine,e15062100}, as it enjoys the GKLS form (cf. Sec.~\ref{sec:introduction}). Interestingly, it is the secular approximation which endows the GME with thermodynamic consistency: The Markovian Redfield equation \eqref{eq:RedfieldCooked}, i.e. the previous step in the derivation of Eq.~\eqref{eq:GKLS}, is known to break positivity \cite{gaspard1999slippage} and caution must be exercised when using it \cite{jeske2015bloch}.		

Coming back to our problem of describing heat transport in the limit of quasi-resonant weakly-coupled nodes, notice that the secular approximation is not problematic when invoked in the derivation of Eq.~\eqref{eq:local_formal}. Indeed, the operators $ \pmb{L}_\alpha^\omega $ may be expanded as a power series in $ k $ in Eq.~\eqref{eq:RedfieldCooked}. At the zeroth order in $ k $, each heat bath would contribute to the right-hand side of Eq.~\eqref{eq:RedfieldCooked} with one non-oscillatory secular term and two fast-rotating non-secular terms at frequencies $ \pm 2\omega_\alpha $. These may be safely averaged out provided that $ \omega_\alpha\gg\lambda^2 $ and regardless of the detuning between the nodes. Consequently, and unlike Eq.~\eqref{eq:global}, the LME should correctly describe the stationary properties of our system when $ k/\omega_c\lesssim\lambda^2 $.

More generally, one can claim that energy transport through an arbitrarily long harmonic chain is correctly captured by a LME within its range of validity; that is, whenever the inter-node couplings are weak. The claim can be made extensive to heat fluxes on spin chains, which were already addressed in Ref.~\cite{PhysRevE.76.031115} via a perturbative master equation relying on `weak internal couplings', precisely in order to bypass the problems created in the GME by the secular approximation.

Finally, let us note that a natural alternative to the LME in our problem would be to incorporate the problematic decay channel of frequency $ \varOmega_+-\varOmega_- $ into the GME, thus arriving to a partial Markovian Redfield equation (cf. \hyperref[app:redfield]{Appendix}). However, scaling up the system in the number of nodes would quickly render this approach too involved to be practical.

\section{Exact non-equilibrium steady state}\label{sec:exact}

The steady state for an all-linear model can also be found \textit{exactly} by solving the corresponding quantum Langevin equations \cite{grabert1984quantum,weiss2008quantum,boyanovsky2017heisenberg}. In this section, we will limit ourselves to outline the procedure to calculate the stationary covariances for our particular problem, while full details on its application to similar settings can be found in e.g. Refs. \cite{Ludwig,PhysRevA.86.012110,PhysRevA.88.042303,PhysRevA.88.012309,PhysRevE.91.062123}.

To begin with, we must mention that the bare frequencies of the nodes need to be shifted so as to compensate for the distortion caused by the system-baths interaction. This eventually allows to recover the correct high temperature limit \cite{weiss2008quantum}. Hence, in the reminder of this section, we shall make the replacement $ \omega^2_\alpha\mapsto\tilde{\omega}_\alpha^2 $, where $ \tilde{\omega}_\alpha^2 \coloneqq \omega_\alpha^2 + \sum_{\mu} g_{\alpha,\mu}^2/(m_{\alpha,\mu}\omega_{\alpha,\mu}^2)$. For our choice of spectral density \eqref{eq:spectraldensity}, the shift amounts simply to $ \pi^{-1}\int_0^{\infty}\,\dif\omega\,J(\omega)/\omega = \lambda^2\Lambda $.

Starting from Eq.~\eqref{eq:hamiltonian}, one may write the Heisenberg equations of motion for all degrees of freedom. Formally solving for $ \pmb{x}_{\alpha,\mu} $ and inserting the result into the equations for $ \pmb{X}_\alpha $ yields the quantum Langevin equations
\begin{align}
\frac{\dif\,^2\pmb{X}_\alpha}{\dif t^2}  + \tilde{\omega}_\alpha^2\pmb{X}_\alpha + k(\pmb{X}_\alpha-\pmb{X}_{\bar{\alpha}}) - \int_{t_0}^\infty\dif s\, \chi_\alpha(t-s)\,\pmb{X}_\alpha(s)= \pmb{F}_\alpha(t) \qquad(\bar{\alpha}\neq\alpha).
\label{eq:QLE}
\end{align}
These are the equations of motion for two coupled harmonic oscillators, each of which is perturbed by the noise $ \pmb{F}_\alpha(t) $ and relaxes according to the dissipation kernel $ \chi_\alpha(t) $. Specifically, these are defined as
\begin{subequations}
\begin{align}
&\pmb{F}_\alpha \coloneqq \sum_{\mu} g_{\alpha,\mu} \big[ \pmb{x}_{\alpha,\mu}(t_0)\cos{\omega_0(t-t_0)} + \frac{\pmb{p}_{\alpha,\mu}(t_0)}{m_{\alpha,\mu}\omega_{\alpha,\mu}}\sin{\omega_{\alpha,\mu}(t-t_0)} \big]\label{eq:fluctuating_force} \\
&\chi_\alpha(t) \coloneqq \sum_{\mu}\frac{g_{\alpha,\mu}^2}{m_{\alpha,\mu}\omega_{\alpha,\mu}}\sin{\omega_{\alpha,\mu}t}\,\varTheta(t) = \frac2\pi\,\varTheta(t)\int_0^\infty\dif\omega\,J(\omega)\sin{\omega t}\label{eq:dissipation_kernel}.
\end{align}
\label{eq:noise_diss}
\end{subequations}

The only assumption that we will make in order to find the exact steady state is, once again, that system and baths are initialized in the factorized initial condition $ \pmb{\rho}_0 = \pmb{\sigma}(t_0)\otimes\pmb{\tau}_c\otimes\pmb{\tau}_h $. We shall also take $ t_0\rightarrow- \infty $. Let us first concentrate on the (stationary)  covariance $ \frac12\langle \{\pmb{X}_\alpha(t),\pmb{X}_{\alpha'}(t)\}_+\rangle $, which may be written in terms of the Fourier transform $ \hat{\pmb{X}}_\alpha(\omega)\coloneqq\int_{-\infty}^{\infty}\dif\,t\,\pmb{X}_\alpha(t)e^{i\omega t} $ as
\begin{equation}
\frac12\langle\{ \pmb{X}_\alpha(t),\pmb{X}_{\alpha'}(t) \}_+\rangle = \frac12\int_{-\infty}^{\infty}\frac{\dif\omega'}{2\pi}\int_{-\infty}^\infty\frac{\dif\omega''}{2\pi}\langle \{\hat{\pmb{X}}_\alpha(\omega'),\hat{\pmb{X}}_{\alpha'}(\omega'')\}_+ \rangle\,e^{-i(\omega' +\omega'')t}.
\label{eq:position-position_formal}
\end{equation}

In turn, $ \hat{\pmb{X}}(\omega)_\alpha $ can be directly found after Fourier-transforming Eqs.~\eqref{eq:QLE}, which yields
\begin{equation}
\begin{pmatrix}
\hat{\pmb{X}}_c \\
\hat{\pmb{X}}_h
\end{pmatrix}
\coloneqq \mathsf{A}^{-1}
\begin{pmatrix}
\hat{\pmb{F}}_c \\
\hat{\pmb{F}}_h
\end{pmatrix} =
\begin{pmatrix}
\tilde{\omega}_c^2-\omega^2+k-\hat{\chi}_c && -k \\
-k && \tilde{\omega}_h^2-\omega^2+k-\hat{\chi}_h
\end{pmatrix}^{-1}
\begin{pmatrix}
\hat{\pmb{F}}_c \\
\hat{\pmb{F}}_h
\end{pmatrix}.
\label{eq:FT_QLE}
\end{equation}

From Eq.~\eqref{eq:fluctuating_force}, one can show that $ \frac12\langle \{\hat{\pmb{F}}_\alpha(\omega'),\hat{\pmb{F}}_{\alpha'}(\omega'')\}_+ \rangle = 2\pi\delta(\omega'+\omega'')\coth{\frac{\omega'}{2T_\alpha}}[J(\omega')\varTheta(\omega')-J(-\omega')\varTheta(-\omega')]\delta_{\alpha,\alpha'}$, where the Dirac delta $ \delta(\cdot) $ is not to be confused with the Kronecker delta $ \delta_{\alpha,\alpha'} $. Consequently, the integral in Eq.~\eqref{eq:position-position_formal} for e.g. $ \alpha = \alpha' = c $ writes as
\begin{equation}
\langle\pmb{X}_c^2\rangle = \int_{-\infty}^{\infty}\frac{\dif\omega}{2\pi}\,\left([\,\mathsf{A}^{-1}(\omega)]_{11}[\,\mathsf{A}^{-1}(-\omega)]_{11}\,\coth{\frac{\omega}{2T_c}}J(\omega) + [\,\mathsf{A}^{-1}(\omega)]_{12}[\,\mathsf{A}^{-1}(-\omega)]_{12}\,\coth{\frac{\omega}{2T_h}}J(\omega)\right),
\label{eq:position-position-cooked}
\end{equation}
where we are exploiting the fact that our $ J(\omega) $ is an odd function. The position-momentum and momentum-momentum covariances are readily obtained as e.g. $ \frac12\langle\{ \pmb{P}_\alpha(t'),\pmb{X}_{\alpha'}(t'') \}_+\rangle = \frac12\int_{-\infty}^{\infty}\frac{\dif\omega'}{2\pi}\int_{-\infty}^{\infty}\frac{\dif\omega''}{2\pi}(-i\omega')\langle \{ \hat{\pmb{X}}_\alpha(\omega'),\hat{\pmb{X}}_{\alpha'}(\omega'') \}_+ \rangle\,e^{-i(\omega't'+\omega''t'')}$.

In order to calculate $ \hat{\chi}_\alpha(\omega) $ it is useful to note that $ \text{Im}\,\hat{\chi}_\alpha(\omega) = J(\omega)\varTheta(\omega)-J(-\omega)\varTheta(-\omega) $, and that $ \text{Re}\,\hat{\chi}_\alpha(\omega) $ and $ \text{Im}\,\hat{\chi}_\alpha(\omega) $ are related through the Kramers-Kronig relation
\begin{equation}
\text{Re}\,\hat{\chi}_\alpha(\omega) = \frac1{\pi}\text{P}\int_{-\infty}^\infty\dif\omega'\frac{\text{Im}\,\hat{\chi}_\alpha(\omega')}{\omega'-\omega},
\label{eq:Kramers-Kronig}
\end{equation}
where $ \text{P} $ indicates Cauchy principal value. For our choice of spectral density $ \hat{\chi}_h(\omega) = \hat{\chi}_c(\omega) = \lambda^2\Lambda^2/(\Lambda-i\omega) $ which, combined with Eqs.~\eqref{eq:spectraldensity}, \eqref{eq:FT_QLE}, and \eqref{eq:position-position-cooked}, allows us to compute all the elements of the exact steady-state covariance matrix $ \mathsf{\Gamma} $. Finally, following Refs.~\cite{PhysRevE.90.042128,PhysRevE.95.012146}, we can cast the exact steady-state heat currents as
\begin{equation}
\dot{\pazocal{Q}}_h = -\dot{\pazocal{Q}}_c = \frac{k}{2}([\mathsf{\Gamma}]_{14}-[\mathsf{\Gamma}]_{23}).
\label{eq:heat_exat}
\end{equation}

Both the steady state covariances and the corresponding heat currents can be seen to perfectly coincide with those obtained from the Markovian Redfield equation derived in the \hyperref[app:redfield]{Appendix}, always provided that the Born-Markov approximation holds.

\section{Discussion}\label{sec:discussion}

\subsection{Steady state and stationary heat currents}\label{subsec:steady-state-heat-currents}

\begin{figure*}
	\includegraphics[width=0.45\linewidth]{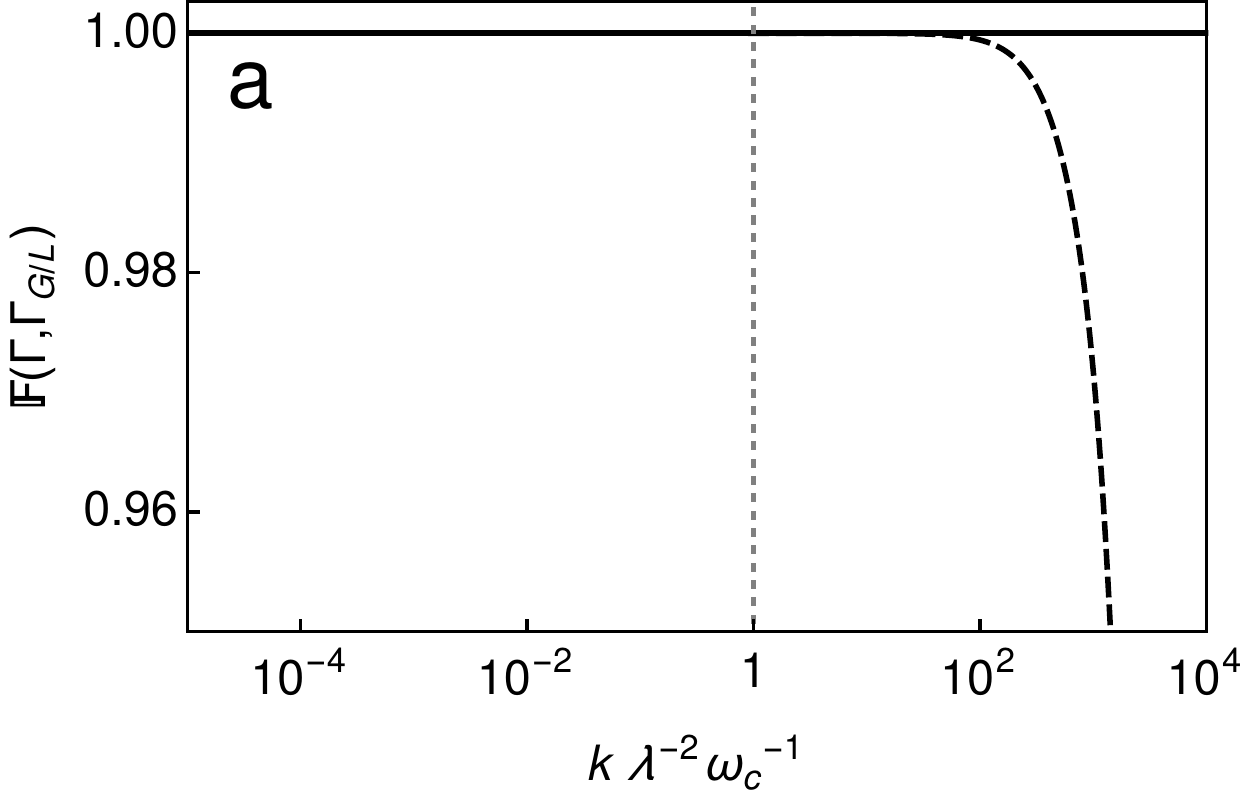}
	\includegraphics[width=0.45\linewidth]{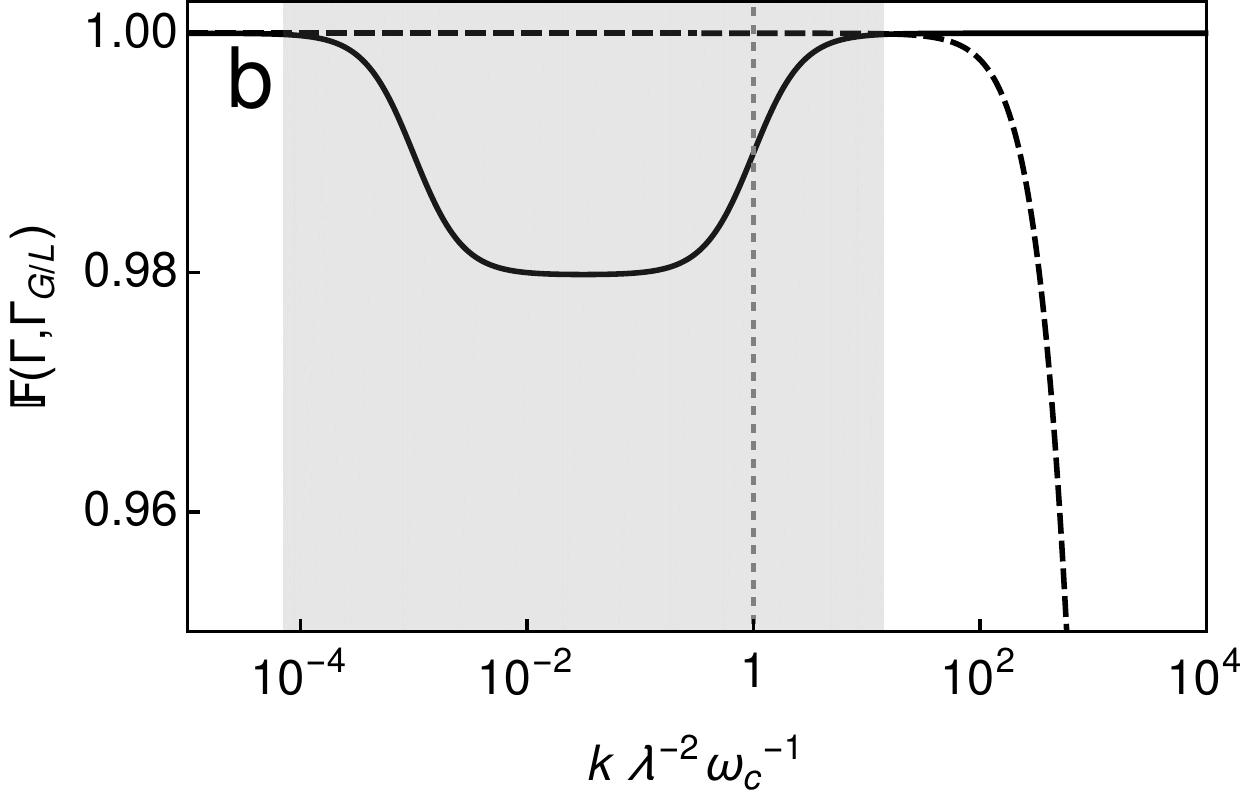}
	\includegraphics[width=0.45\linewidth]{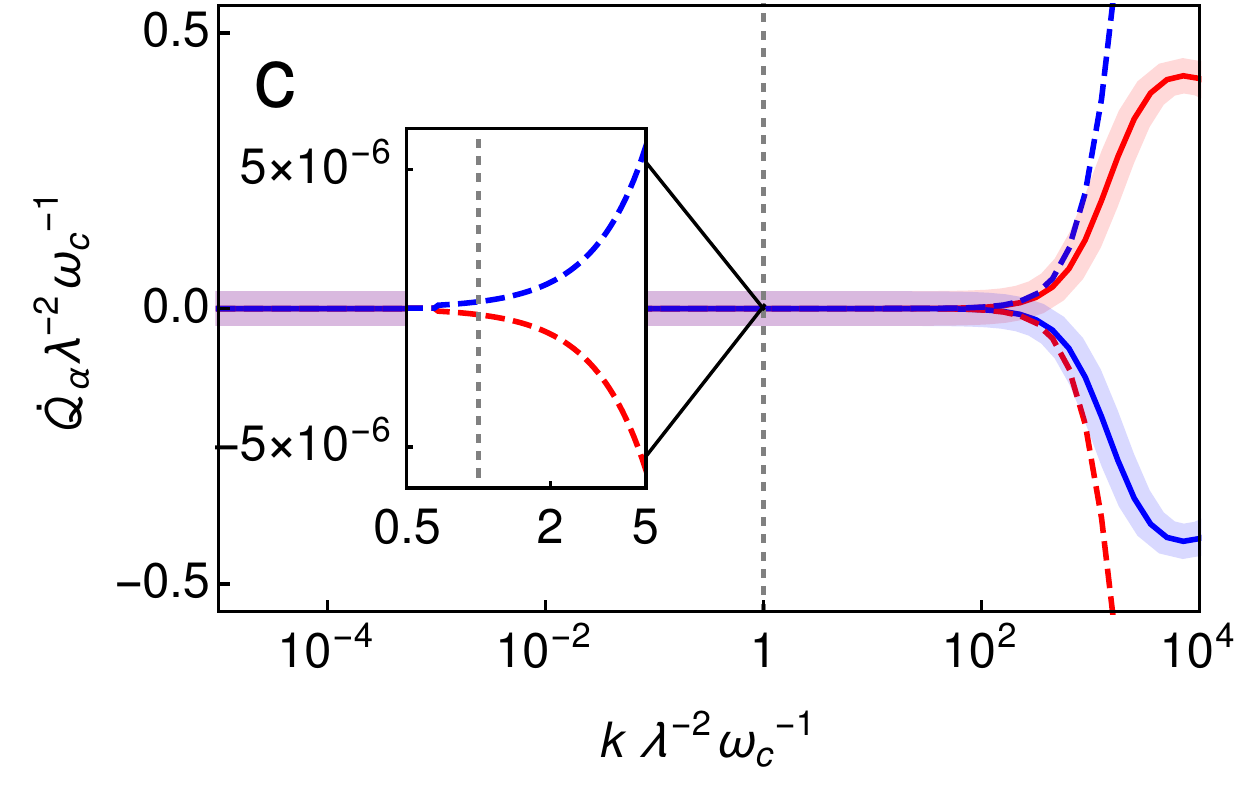}
	\includegraphics[width=0.45\linewidth]{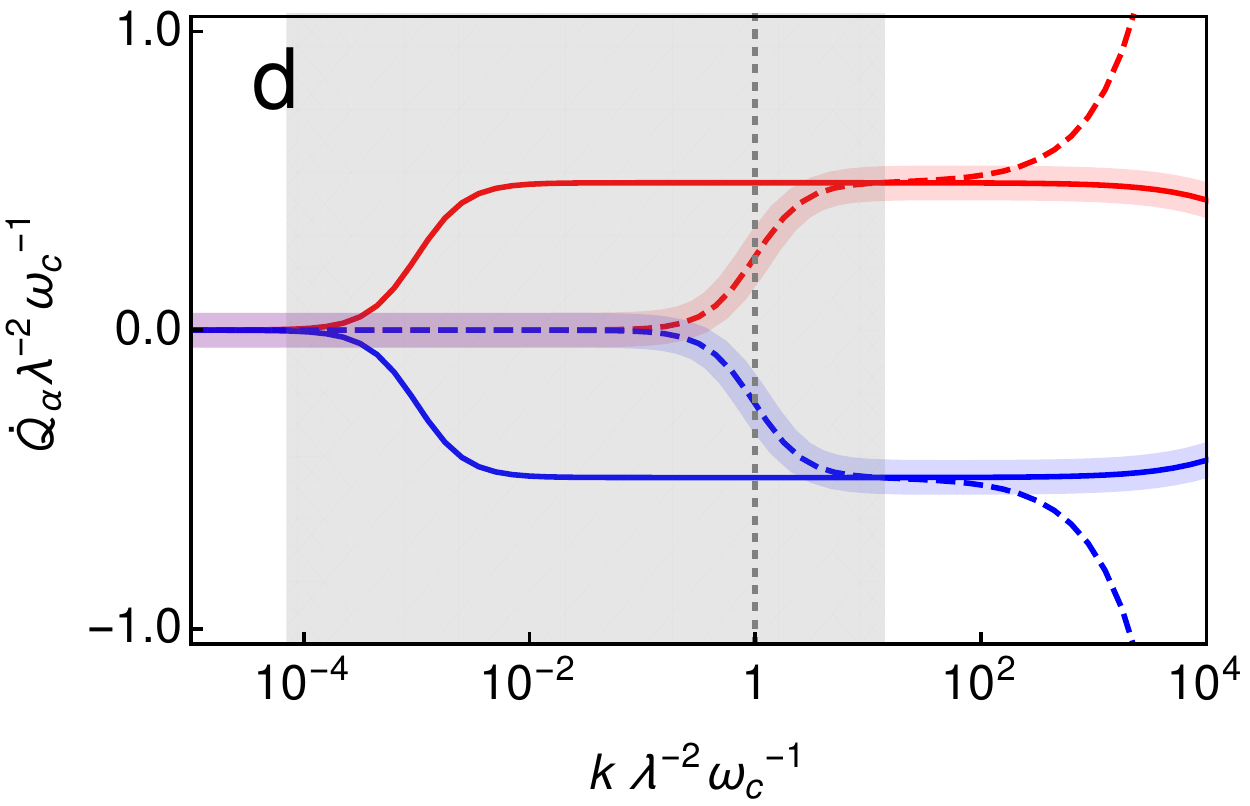}
	\caption{(color online) \textbf{(top row)} Uhlmann fidelity $ \mathbb{F} $ between the exact steady state $ \mathsf{\Gamma} $ and the approximations $ \mathsf{\Gamma}^\text{G} $ and $ \mathsf{\Gamma}^\text{L} $ calculated wihtin the global (solid) and local (dashed) approach, as a function of the coupling $ k $ at fixed dissipation strength $ \lambda^2 = 10^{-3} $. In \textbf{(a)} frequencies and temperatures were set to $ \omega_h = 2 $, $ \omega_c = 1 $, $ T_h = 3 $, and $ T_c = 2 $, so that $ \delta_\omega \gg \lambda^2 $ and the secular approximation is justified. Hence, the global GKLS equation is in perfect agreement with the exact result. The LME starts to break down around $ k\sim 0.1 $, i.e. when the inter-node coupling becomes comparable to the node frequencies. In \textbf{(b)} the nodes are quasi-resonant ($ \omega_c = 1 $ and $ \delta_\omega^2 = 2\times10^{-6} $), while the temperatures are the same as in \textbf{(a)}. Due to the breakdown of the secular approximation, the global GKLS equation becomes unreliable. The shaded grey area corresponds to $ 1-\mathbb{F}(\mathsf{\Gamma},\mathsf{\Gamma}^\text{G})\geq 10^{-4} $. In contrast, the LME remains accurate in that regime of parameters. \textbf{(bottom row)} Stationary incoming heat currents from the hot (red) and cold (blue) baths, as given by the global (thin solid), local (dashed), and exact (thick transparent) approaches. The parameters in \textbf{(c)} are the same as in \textbf{(a)}. As it can be seen, the LME violates the second law of thermodynamics predicting \textit{reversed} heat currents for all $ k $. Finally, the parameters in \textbf{(d)} are the same as in \textbf{(b)}. In the shaded grey region, in which the secular approximation breaks down, the GME greatly overestimates the magnitude of the steady-state heat currents, while the LME perfectly follows the exact result. For all four plots $ \Lambda = 10^3 $. Recall that we work in units of $ m_c=m_h=\hbar=k_B=1 $.}
	\label{fig1}
\end{figure*}

In this section we will compare the steady states and the stationary heat currents predicted by the global, local, and exact approaches. We shall be especially interested in setting up the wire with quasi-resonant nodes ($ \delta_\omega \ll \lambda^2 $) so as to confirm our intuition that the LME can succeed in describing the system when the GME breaks down (cf. Sec.~\ref{subsec:comment}).

In order to compare states we will make use of the Uhlmann fidelity, defined as $ \mathbb{F}(\pmb{\rho}_1,\pmb{\rho}_2) \coloneqq (\tr\sqrt{\sqrt{\pmb{\rho}_1}\pmb{\rho}_2\sqrt{\pmb{\rho}_1}})^2 $ for arbitrary $ \pmb{\rho}_1 $ and $ \pmb{\rho}_2 $ \cite{nielsen2000}. In the case of two-mode Gaussian states with covariance matrices $ \mathsf{\Gamma}_1 $ and $ \mathsf{\Gamma}_2 $ and vanishing first order moments the fidelity can be cast as \cite{PhysRevA.86.022340}
\begin{equation}
\mathbb{F}(\mathsf{\Gamma}_1,\mathsf{\Gamma}_2) = \left[\left(\sqrt{\mathit{b}}+\sqrt{\mathit{c}}\right)-\sqrt{\left(\sqrt{\mathit{b}}+\sqrt{\mathit{c}}\right)^2-\mathit{a}}\right]^{-1},
\label{eq:fidelity_2mode}
\end{equation}
where $ \mathit{a}\coloneqq\det{(\mathsf{\Gamma}_1+\mathsf{\Gamma}_2)} $, $ \mathit{b}\coloneqq 2^4\det{[(\mathsf{J}\mathsf{\Gamma}_1)(\mathsf{J}\mathsf{\Gamma}_2)-\mathbb{I}/4]} $, $ \mathit{c}\coloneqq 2^4\det{(\mathsf{\Gamma}_1+i\mathsf{J}/2)}\det{(\mathsf{\Gamma}_2+i\mathsf{J}/2)} $, and $ \mathsf{J}_{kl} \coloneqq -i[\pmb{R}_k,\pmb{R}_l] $.

As shown in Fig.~\hyperref[fig1]{1(a)}, whenever the detuning is large compared with the dissipation strength, both LME and GME are in perfect agreement with the exact solution for most parameters. The local approach only starts to break down when the coupling $ k $ becomes comparable or larger than the node frequencies (i.e. $ k/\omega_\alpha \gtrsim 0.1 \omega_c $, where the extra $ \omega_c $ has been merely added for dimensional consistency), whereas the global master equation remains correct.

On the contrary, if the detuning is set to $ \delta_\omega\ll\lambda^2 $, the steady state of the GME can be seen to disagree with the exact solution when the inter-node coupling $ k/\omega_c $ approaches or falls below the dissipation strength $ \lambda^2 $ [cf. Fig.~\hyperref[fig2]{2(b)}]. Recall that this is entirely due to elimination of the non-secular decay channel at frequency $ \varOmega_+-\varOmega_- $ (cf. Sec.~\ref{subsec:global}). Importantly, the LME is still valid so long as $ k/\omega_c \ll \omega_c $, \textit{regardless} of the breakdown of the secular approximation. Eventually, as $ k $ decreases further, the nodes effectively decouple, and the GME correctly predicts a steady state made up of two uncorrelated thermal modes.

One can also make use of Eqs.~\eqref{eq:heat_GME}, \eqref{eq:heat_LME}, and \eqref{eq:heat_exat} to compare the steady-state heat currents. Once again, under large detuning $ \delta_\omega $, both the local and global approach are in good agreement with the exact solution (vanishingly small heat currents), except for when the inter-node coupling becomes comparable to the node frequencies, which invalidates the LME. Interestingly, in Fig.~\hyperref[fig1]{1(c)} we can see that the local approach does indeed violate the second law of thermodynamics by predicting heat transport against the temperature gradient (i.e. $ \dot{\pazocal{Q}}_h = - \dot{\pazocal{Q}}_c < 0 $) for any $ k $ \cite{levy2014local}. The magnitude of this violation, however, loosely falls within the `error bars' $ \pazocal{O}(\lambda^2 k) $ \cite{trushechkin2016perturbative} of the LME.

On the other hand, Fig.~\hyperref[fig1]{1(d)} shows again a situation in which $ \delta_\omega\ll\lambda^2 $. Remarkably, we observe that the global approach largely overestimates the magnitude of the steady-state heat currents where $ \mathbb{F}(\mathsf{\Gamma},\mathsf{\Gamma}^\text{G}) $ falls below $ 1 $ (i.e. in the grey area). The LME, however, yields a quantitatively good estimate in all the range of parameters for which it is valid.

We have thus illustrated that the breakdown of the secular approximation may render the predictions of the global master equation \textit{qualitatively} wrong, while the local approach, in spite of its thermodynamic inconsistency, proves to be an accurate working tool within its range of applicability.

\subsection{Steady-state correlations}\label{subsec:correlations}

\begin{figure*}
	\includegraphics[width=0.45\linewidth]{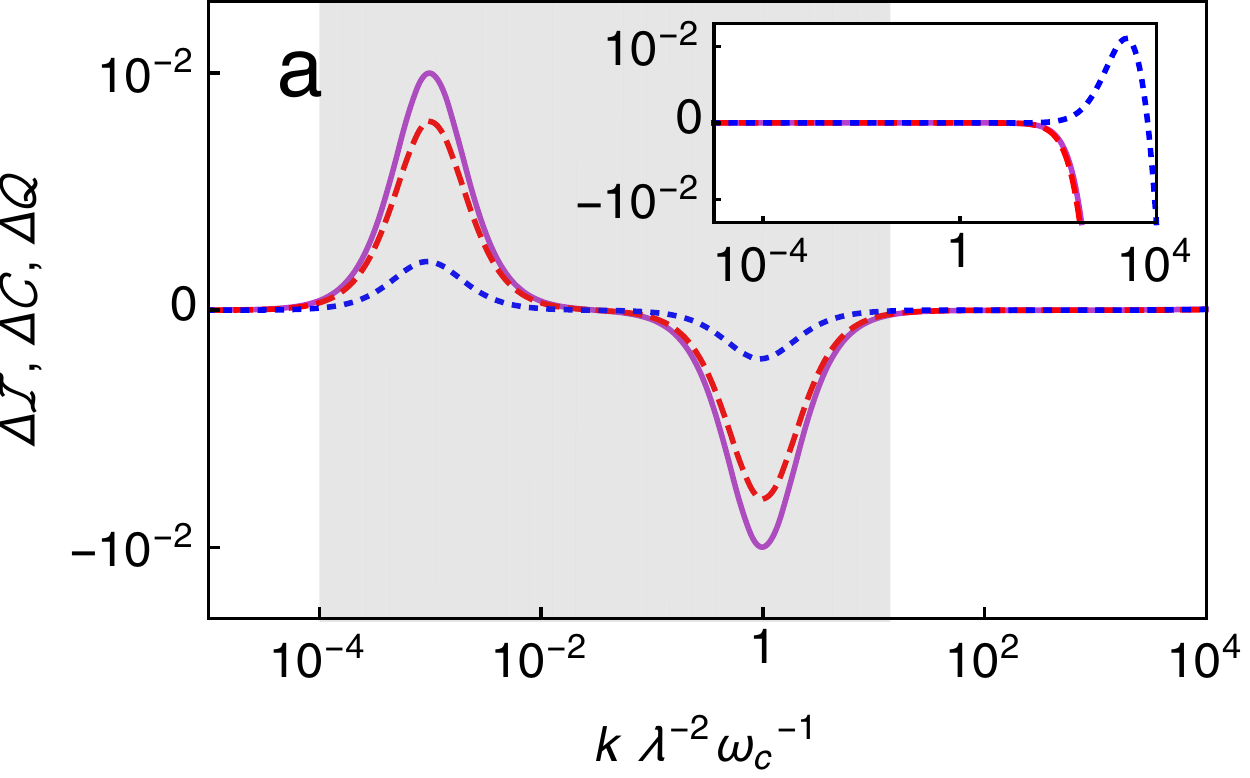}
	\includegraphics[width=0.27\linewidth]{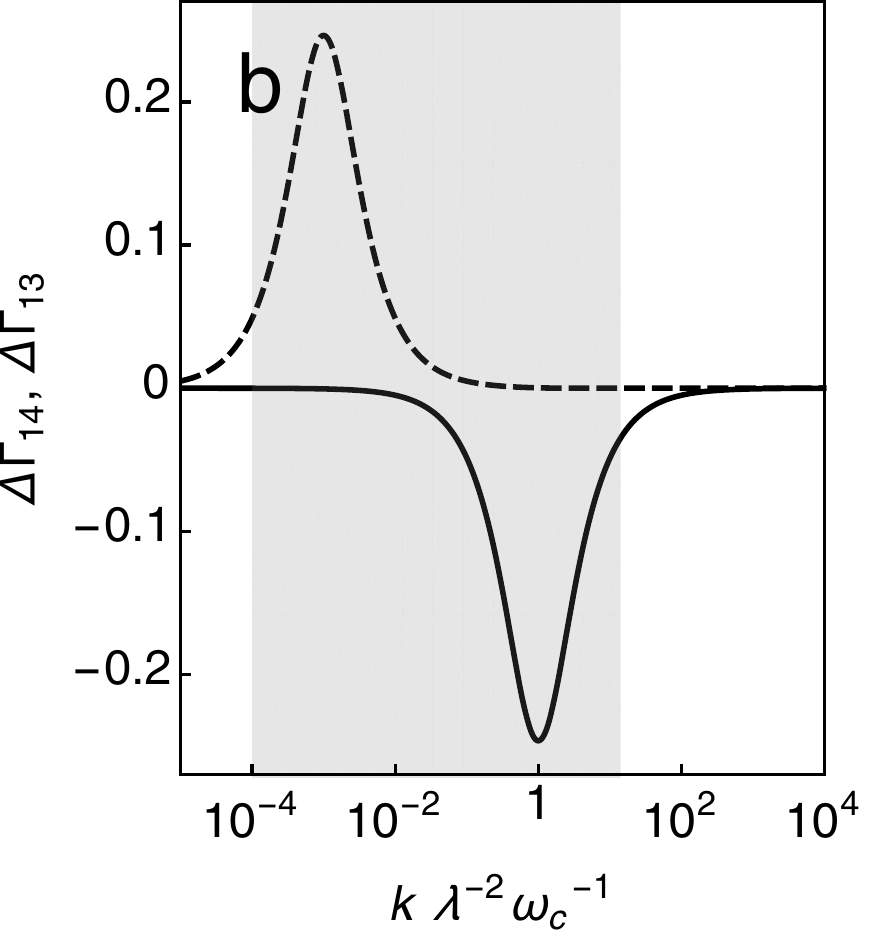}
	\includegraphics[width=0.26\linewidth]{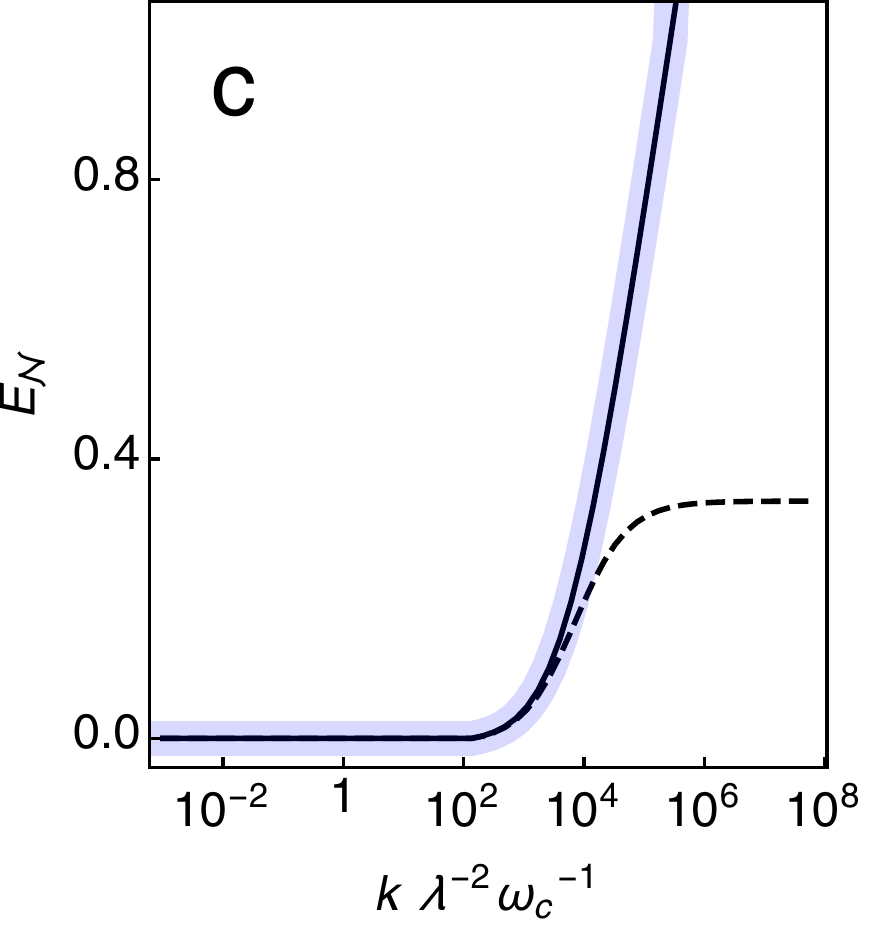}
	\caption{(color online) \textbf{(a)} Excess quantum mutual information $\Delta\pazocal{I}\coloneqq\pazocal{I}(\mathsf{\Gamma}^\text{G})-\pazocal{I}(\mathsf{\Gamma}) $ (solid purple), classical correlations $ \Delta\pazocal{C}^\leftarrow\coloneqq\pazocal{C}^\leftarrow(\mathsf{\Gamma}^\text{G})-\pazocal{C}^\leftarrow(\mathsf{\Gamma}) $ (dashed red), and quantum correlations $ \Delta\pazocal{Q}^\leftarrow=\Delta\pazocal{I}-\Delta\pazocal{C}^\leftarrow $ (dotted blue), as follows from the comparison of the global approach with the exact solution (see main text for definitions). The inset reproduces the main plot, benchmarking instead the local approach against the exact result. The parameters are the same as in Figs.~\hyperref[fig1]{1(b)} and \hyperref[fig1]{1(d)}, i.e. we work with small detuning. While the LME faithfully captures inter-node correlations in all its range of validity, the GME may both underestimate or overestimate them. In \textbf{(b)} the quantities $\Delta\mathsf{\Gamma}_{14}\coloneqq [\mathsf{\Gamma}^\text{G}]_{14}-[\mathsf{\Gamma}]_{14}$ (solid) and $ \Delta\mathsf{\Gamma}_{13}\coloneqq[\mathsf{\Gamma}^\text{G}]_{13}-[\mathsf{\Gamma}]_{13} $ (dashed) are plotted for the same parameters as in (a). We see that the failure of the global approach to correctly assess these covariances ($ \langle\pmb{x}_c\,\pmb{p}_h\rangle = -\langle\pmb{p}_c\,\pmb{x}_h\rangle $ and $ \langle\pmb{x}_c\,\pmb{x}_h\rangle $) explains the peaks in (a). In \textbf{(c)} we plot the steady state inter-node entanglement, as quantified by the logarithmic negativity $ E_\pazocal{N} $ within the global (solid black), local (dashed black), and exact (thick transparent blue) approaches. The observation of non-vanishing asymptotic entanglement requires large ratios $ \omega/T_\alpha $ and very large coupling strengths $ k $. Unfortunately, this prevents entanglement from being observed in the problematic region $ k/\omega_c\lesssim\lambda^2 $. Interestingly, the LME predicts a saturation of $ E_\pazocal{N} $ for large $ k $, which is anyway far beyond its range of applicability. In (c) $ \omega_c=\omega_h=10 $, $ T_c = 1 $, $ T_h = 2 $, $ \lambda^2 = 10^{-3} $, and $ \Lambda = 10^3 $.}
	\label{fig2}
\end{figure*}

As we shall now see, the GME also fails qualitatively in assessing the node-node correlations (both classical and quantum) when the secular approximation breaks down. This is not the case for the LME.

We measure the total correlations between the `cold' and `hot' nodes of the wire by means of the quantum mutual information $ \pazocal{I}(\pmb{\sigma}_{c\,h})\coloneqq S(\pmb{\sigma}_c) + S(\pmb{\sigma}_h) - S(\pmb{\sigma}_{c\,h}) $, where $ S(\pmb{\varrho}) = -\tr{\{ \pmb{\varrho} \log{\pmb{\varrho}}\}}$ is the von Neumann entropy and $ \pmb{\sigma}_\alpha \coloneqq \tr_{\bar{\alpha}}\,\pmb{\sigma}_{c\,h} $ stands for the reduced state of node $ \alpha $ (the subindices ` $ c $' and `$ h $' are added to the starionary state of the wire to emphasize its bipartite nature). The von Neumann entropy of an $ n $-mode Gaussian state can be written as \cite{PhysRevA.63.032312}
\begin{equation}
S(\mathsf{\Gamma}) = \sum\nolimits_{j=1}^n\left(\frac{2\nu_j + 1}{2}\,\log{\frac{2\nu_j+1}{2}} + \frac{2\nu_j - 1}{2}\,\log{\frac{2\nu_j-1}{2}} \right),
\end{equation}
where the $ \nu_j $ are the $ n $ \textit{symplectic eigenvalues} of the generic $ 2n\times 2n $ covariance matrix $ \mathsf{\Gamma} $. These can be obtained form the spectrum $ \{\pm i \nu_1,\cdots,\pm i \nu_n \}$ of $ \mathsf{J}^{-1}\mathsf{\Gamma} $. For $ \mathsf{\Gamma} $ to be physical, the symplectic spectrum must satisfy	 $ \nu_j\geq\frac12 $. In our case, note that e.g. the single-mode covariance matrix $ \mathsf{\Gamma}_c $ results from retaining only the first two rows and columns of the two-mode covariance matrix of the full system, i.e. those related to the `cold quadratures' $ \{\pmb{x}_c,\pmb{p}_c\} $.

As we can see from Fig.~\hyperref[fig2]{2(a)} the inter-node correlations can be both overestimated and underestimated by the global master equation whenever the secular approximation fails. In contrast, the LME assesses $ \pazocal{I} $ faithfully. Note from Eq.~\eqref{eq:cov_global} that the stationary covariances $ \langle\pmb{x}_c\,\pmb{p}_h\rangle $ and $ \langle \pmb{p}_c\,\pmb{x}_h \rangle $ (i.e. $ \mathsf{\Gamma}_{14} $ and $ \mathsf{\Gamma}_{23} $) are neglected in the global approach. Indeed, it is easy to see from the corresponding Markovian Redfield equation (cf. \hyperref[app:redfield]{Appendix}) that these covariances are related to the excluded non-secular term at frequency $ \varOmega_+-\varOmega_- $. From Fig.~\hyperref[fig2]{2(b)} we observe that the \textit{deficit} in total quantum correlations predicted by the GME around $ k/\omega_c\simeq\lambda^2 $ in Fig.~\hyperref[fig2]{2(a)} is precisely due to the fact that $ [\mathsf{\Gamma}^\text{G}]_{14} = [\mathsf{\Gamma}^\text{G}]_{23} = [\mathsf{\Gamma}^\text{G}]_{32} = [\mathsf{\Gamma}^\text{G}]_{41} = 0 $. Notice, comparing again Figs.~\hyperref[fig2]{2(a)} and \hyperref[fig2]{2(b)}, that the peak in the total correlations at lower $ k $ is explained by the fact that the GME overestimates $ \langle\pmb{x}_c\,\pmb{x}_h\rangle $; once again, within the region in which the secular approximation breaks down.

It is possible to split the total correlations into a \textit{quantum} and a \textit{classical} share (blue dotted and red dashed lines in Fig.~\hyperref[fig2]{2(a)}, respectively). We will say that a bipartite quantum state $ \pmb{\varrho}_{AB} $ has quantum correlations with respect to $ B $ if there exists no local measurement on $ B $ that leaves the marginal of $ A $ unperturbed. This notion of \textit{quantumness} of correlations is captured by the discord $ \pazocal{Q}^\leftarrow(\pmb{\varrho}_{AB})\coloneqq S(\pmb{\varrho}_B)-[S(\pmb{\varrho}_{AB})- \inf_{\{\pmb{\Pi}_j^B \}}\sum_j p_j S(\pmb{\varrho}_{A\vert j})] $ \cite{olliver20011,henderson20011}. Given a complete set of projectors $ \{\pmb{\Pi}^B_j \} $ on $ B $, $ \pmb{\varrho}_{A\vert j}\coloneqq\tr_B\{\pmb{\Pi}_j^B\pmb{\varrho}_{AB}\pmb{\Pi}_j^B\} $ denotes the post-measurement marginal of $ A $ conditioned on the outcome $ j $, occuring with probability $ p_j = \tr\{ \pmb{\Pi}_j^B\,\pmb{\varrho}_{AB} \} $. Note that discord is not symmetric, i.e. the quantumness of correlations as revealed by measurements on $ B $ need not coincide with the quantumness of correlations as revealed by measurements on $ A $.

Note as well that, due to the explicit minimization over all local measurements on $ B $, the evaluation of $ \pazocal{Q}^\leftarrow $ is often very challenging. Luckily, restricting the optimization to the set of Gaussian positive operator valued measurements makes it possible to obtain a closed formula for two-mode Gaussian states (see Ref.~\cite{adesso2010quantum,PhysRevLett.105.020503,PhysRevLett.113.140405} for full details). The difference between the total correlations and the quantum discord is referred-to as classical correlations $ \pazocal{C}^\leftarrow(\pmb{\varrho}_{AB})\coloneqq \pazocal{I}(\pmb{\varrho}_{AB}) - \pazocal{Q}^\leftarrow(\pmb{\varrho}_{AB}) $. As shown in Fig.~\hyperref[fig2]{2(a)}, both quantum and classical correlations behave very similarly to the mutual information within the global approach. This is not the case, however, for the LME [cf. inset in Fig.~\hyperref[fig2]{2(a)}]: at large coupling strengths (i.e. beyond its range of validity) the local approach may overestimate the amount of quantum correlations present between the nodes, although at sufficiently large couplings, all correlations are largely underestimated.

Finally, we may want to look at the inter-node \textit{entanglement} \cite{RevModPhys.81.865}. Entanglement is a somewhat stronger form of quantum correlation since a state can display non-zero discord and yet be unentangled, but not the other way around. In the case of two-mode Gaussian states, quantum entanglement can be gauged by the logarithmic negativity $ E_\pazocal{N} $, which writes as \cite{LogarithmicNegativity,PhysRevLett.95.090503}
\begin{equation}
E_\pazocal{N}(\mathsf{\Gamma})\coloneqq\sum_j \max{\{0,-\log{(2\tilde{\nu}_j)}\}},
\label{eq:logarithmic_negativity}
\end{equation}
where $ \tilde{\nu}_j $ are the symplectic eigenvalues of the \textit{partially-transposed} covariance matrix $ \tilde{\mathsf{\Gamma}} $. This is obtained from $ \mathsf{\Gamma} $ by simply changing the sign of all covariances involving e.g. the momentum $ \pmb{p}_c $ and either of the `hot' quadratures.

The buildup of steady-state entanglement requires much larger inter-node coupling $ k $ and large ratios $ \omega_\alpha/T_\alpha $ as shown in Fig.~\hyperref[fig2]{2(c)}. While there is no reason for the GME not to accurately capture the entanglement as $ k\rightarrow\infty $, the LME wrongly predicts a saturation in the stationary logarithmic negativity in that limit. One can obtain the correct scaling of entanglement at strong coupling from the GME which, for resonant nodes, simplifies to
\begin{equation}
E_\pazocal{N}(\mathsf{\Gamma})\mathrel{\mathop{\rightarrow}_{k\rightarrow\infty}}\frac14\log\frac{2k(1-e^{\omega/T_c})^2(1-e^{\omega/T_h})^2}{(1-e^{2\omega/\bar{T}})^2\omega^2}\qquad\text{with}\qquad\bar{T}\coloneqq\big(\frac{T_c^{-1}+T_h^{-1}}{2}\big)^{-1}.
\label{eq:strong_k_lognegativity}
\end{equation}

\section{Conclusions}\label{sec:conclusions}

We have studied a simple model for heat transport between two heat baths at different temperatures when weakly connected through a two-node quantum wire. Due to the weak dissipative wire-baths coupling, it is possible to address the problem via second-order Markovian quantum master equations. In particular, we consistently derived the GKLS master equation via a \textit{global} treatment of dissipation, and found its steady state, the stationary heat currents through the wire, and the asymptotic inter-node quantum and classical correlations. For comparison, we adopted the popular \textit{local} approach, which addresses dissipation on each node individually (i.e. ignoring the effects of the inter-node coupling). Since our model is linear, its steady state can be obtained \textit{exactly} by resorting to quantum Langevin equations. This provided us with means to quantitatively compare the performance of the global and the local approaches.

As expected, we found that the local approach is only valid when the internal coupling between the nodes of the wire is weak. Furthermore, as previously noted, we observed that the local approach does break the second law of thermodynamics \cite{levy2014local}, although any violations can be bounded with suitably-defined error bars within its range of applicability \cite{trushechkin2016perturbative}.

Interestingly, our setup allows us to consider very weak internal couplings, comparable with the dissipation strength. In this regime, the crucial secular approximation breaks down if, in addition, the nodes are nearly resonant. As a result, the predictions of the global master GME become qualitatively wrong---the magnitude of the stationary heat currents is largely overestimated, and key features of the correlation-sharing structure are not captured by the GME. On the contrary, the LME does accurately describe the stationary properties of the wire. This agrees with previous observations on the complementarity of GME and LME when describing dynamics \cite{rivas2010markovian}. More generally, the usage of the local approach in the treatment of heat transport through arbitrarily long harmonic or spin chains \cite{PhysRevE.76.031115} may be justified provided that the internal couplings are weak enough, and always keeping in mind that the predictions of the LME should by accompanied by the corresponding error estimates \cite{trushechkin2016perturbative}.

In spite of these encouraging observations, the local approach should not be used lightly, especially in quantum thermodynamics. Even though the LME may be an excellent working tool that even outperforms the canonical global GKLS master equation in certain regimes, it might as well lead to qualitatively wrong conclusions, \textit{a priori} within its range of applicability. For instance, it has been shown that a local modelling of quantum thermodynamic cycles completely fails to account for heat leaks and internal dissipation effects \cite{PhysRevE.87.042131,PhysRevE.92.032136} that can become dominant in the operation of the device in question. As a result, e.g. intrinsically irreversible models may be wrongly classified as \textit{endoreversible}. This is a reminder that perturbative equations of motion for open quantum systems must always be handled with care.

\textbf{\textsc{Note added:}} During the preparation of this manuscript we became aware of the related work by Patrick P. Hofer \textit{et al.} \cite{hofer2017markovian}, where local and global approach are compared in a quantum heat engine model.   

\begin{acknowledgements}
The authors gratefully acknowledge A. Levy, Nahuel Freitas, and Karen V. Hovhannisyan for helpful comments. This project was funded by the Spanish MECD (FPU14/06222), the Spanish MINECO (FIS2013-41352-P), the European Research Council under the StG GQCOP (Grant No. 637352), and the COST Action MP1209: ``Thermodynamics in the quantum regime''. 	
\end{acknowledgements}	

\appendix*

\section{The \textit{partial} Markovian Redfield master equation}\label{app:redfield}

In order to compensate for the deficiencies of the GME one may simply take into consideration the problematic non-secular term corresponding to the $ \varOmega_+-\varOmega_- $ channel. Eqs.~\eqref{eq:RedfieldCooked} and \eqref{eq:adjoint_GKLS} would then need to be combined as
\begin{align}
\begin{split}
\frac{\dif \pmb{O}}{\dif t} \simeq i[\pmb{H}_S,\pmb{O}] &+ \sum_{\alpha\in\{c,h\}}\sum_{\omega \in\{\pm\varOmega_\pm\}}\gamma_\alpha(\omega)\left( {\pmb{L}_\alpha^\omega}^\dagger\pmb{O}\,\pmb{L}_\alpha^\omega -\frac12\{ {\pmb{L}_\alpha^\omega}^\dagger\pmb{L}_\alpha^\omega,\pmb{O} \}_+ \right) \\
& + \frac12\sum_{\alpha\in\{ c,h \}} \gamma_\alpha(\varOmega_+)\left( {\pmb{L}_\alpha^{\varOmega_-}}^\dagger\pmb{O}\,\pmb{L}_\alpha^{\varOmega_+} - \pmb{O}\,{\pmb{L}_\alpha^{\varOmega_-}}^\dagger\pmb{L}_\alpha^{\varOmega_+} + {\pmb{L}_\alpha^{\varOmega_+}}^\dagger\pmb{O}\,\pmb{L}_\alpha^{\varOmega_-} - {\pmb{L}_\alpha^{\varOmega_+}}^\dagger\pmb{L}_\alpha^{\varOmega_-}\pmb{O}\right)\\
& + \frac12\sum_{\alpha\in\{ c,h \}} \gamma_\alpha(-\varOmega_+)\left( {\pmb{L}_\alpha^{\varOmega_-}}\pmb{O}\,{\pmb{L}_\alpha^{\varOmega_+}}^\dagger - \pmb{O}\,{\pmb{L}_\alpha^{\varOmega_-}}{\pmb{L}_\alpha^{\varOmega_+}}^\dagger + {\pmb{L}_\alpha^{\varOmega_+}}\pmb{O}\,{\pmb{L}_\alpha^{\varOmega_-}}^\dagger - {\pmb{L}_\alpha^{\varOmega_+}}{\pmb{L}_\alpha^{\varOmega_-}}^\dagger\pmb{O}\right)\\
& + \frac12\sum_{\alpha\in\{ c,h \}} \gamma_\alpha(\varOmega_-)\left( {\pmb{L}_\alpha^{\varOmega_+}}^\dagger\pmb{O}\,\pmb{L}_\alpha^{\varOmega_-} - \pmb{O}\,{\pmb{L}_\alpha^{\varOmega_+}}^\dagger\pmb{L}_\alpha^{\varOmega_-} + {\pmb{L}_\alpha^{\varOmega_-}}^\dagger\pmb{O}\,\pmb{L}_\alpha^{\varOmega_+} - {\pmb{L}_\alpha^{\varOmega_-}}^\dagger\pmb{L}_\alpha^{\varOmega_+}\pmb{O}\right) \\
& + \frac12\sum_{\alpha\in\{ c,h \}} \gamma_\alpha(-\varOmega_-)\left( {\pmb{L}_\alpha^{\varOmega_+}}\pmb{O}\,{\pmb{L}_\alpha^{\varOmega_-}}^\dagger - \pmb{O}\,{\pmb{L}_\alpha^{\varOmega_+}}{\pmb{L}_\alpha^{\varOmega_-}}^\dagger + {\pmb{L}_\alpha^{\varOmega_-}}\pmb{O}\,{\pmb{L}_\alpha^{\varOmega_+}}^\dagger - {\pmb{L}_\alpha^{\varOmega_-}}{\pmb{L}_\alpha^{\varOmega_+}}^\dagger\pmb{O}\right),
\label{eq:partial_redfield}
\end{split}
\end{align}
where the operators $ \pmb{L}_\alpha^\omega $ are those defined in Sec.~\ref{subsec:global}.

In principle, a full set of $ 10 $ dynamical variables would be necessary to obtain all steady-state covariances. We shall choose $ \pmb{D}_{\pm\pm}\coloneqq i(\pmb{a}_\pm^\dagger\pmb{a}_{\pm}^\dagger-\pmb{a}_\pm\pmb{a}_{\pm}) $, $ \pmb{S}_{\pm\pm}\coloneqq \pmb{a}_\pm^\dagger\pmb{a}_{\pm}^\dagger + \pmb{a}_\pm\pmb{a}_{\pm} $, $ \pmb{D}_{+-}\coloneqq i (\pmb{a}_+^\dagger\pmb{a}_-^\dagger-\pmb{a}_+\pmb{a}_-) $, $ \pmb{S}_{+-}\coloneqq \pmb{a}_+^\dagger\pmb{a}_-^\dagger + \pmb{a}_+\pmb{a}_- $, $ \pmb{d}_{+-}\coloneqq i (\pmb{a}_+^\dagger\pmb{a}_- -\pmb{a}_+\pmb{a}_-^\dagger) $, $ \pmb{s}_{+-}\coloneqq \pmb{a}_+^\dagger\pmb{a}_- + \pmb{a}_+\pmb{a}_-^\dagger $, and $ \pmb{n}_\pm\coloneqq\pmb{a}_\pm^\dagger\pmb{a}_\pm $. As it turns out, the stationary averages of the first six variables vanish (i.e. $ \langle\pmb{D}_{\pm\pm}\rangle = \langle\pmb{S}_{\pm\pm}\rangle = \langle \pmb{D}_{+-} \rangle = \langle\pmb{S}_{+-}\rangle = 0 $), so that we are left with only four relevant observables. The corresponding equations of motion write as $ \dif\vec{\pmb{y}}/\dif t = \mathsf{B}\,\vec{\pmb{y}} + \mathsf{b} $, where $ \vec{\pmb{y}} = (\pmb{n}_+,\pmb{n}_-,\pmb{d}_{+-},\pmb{s}_{+-})^\mathsf{T} $, the non-zero elements of $ \mathsf{b} $ are given by
\begin{subequations}
	\begin{align}
	&[\mathsf{b}]_{1} = W^c_{-\varOmega_+} + W^h_{-\varOmega_+} \\
	&[\mathsf{b}]_{2} = W^c_{-\varOmega_-} + W^h_{-\varOmega_-} \\
	&[\mathsf{b}]_{4} = \sqrt{\frac{\varOmega_+}{\varOmega_-}}(W^c_{-\varOmega_+}\tan{\vartheta} - W^h_{-\varOmega_+}\cot{\vartheta} )+\sqrt{\frac{\varOmega_-}{\varOmega_+}}(W^c_{-\varOmega_-}\cot{\vartheta} - W^h_{-\varOmega_-}\tan{\vartheta}),
	\end{align}
	\label{eq:redfield_vector}
\end{subequations}
and the coefficients of the matrix $ \mathsf{B} $ read
\begin{subequations}
	\begin{align}
	&[\mathsf{B}]_{11} = W^c_{-\varOmega_+} + W^h_{-\varOmega_+} - W^c_{\varOmega_+} - W^h_{\varOmega_+} \\
	&[\mathsf{B}]_{14} = \frac12 [\mathsf{B}]_{42} = \frac{1}{2}\sqrt{\frac{\varOmega_-}{\varOmega_+}} ([W^c_{-\varOmega_-} - W^c_{\varOmega_-}]\cot{\vartheta} - [W^h_{-\varOmega_-} - W^h_{\varOmega_-}]\tan{\vartheta})\\
	&[\mathsf{B}]_{22} = W^c_{-\varOmega_-} + W^h_{-\varOmega_-} - W^c_{\varOmega_-} - W^h_{\varOmega_-} \\
	&[\mathsf{B}]_{24} = \frac12 [\mathsf{B}]_{41} = \frac{1}{2}\sqrt{\frac{\varOmega_+}{\varOmega_-}}([W^c_{-\varOmega_+} - W^c_{\varOmega_+}]\tan{\vartheta} - [W^h_{-\varOmega_+} - W^h_{\varOmega_+}]\cot{\vartheta}) \\
	&[\mathsf{B}]_{33} = [\mathsf{B}]_{44} = \frac{1}{2}(W^c_{-\varOmega_-} + W^c_{-\varOmega_+} + W^h_{-\varOmega_-} + W^h_{-\varOmega_+} - W^c_{\varOmega_-} - W^c_{\varOmega_+} - W^h_{\varOmega_-} - W^h_{\varOmega_+}) \\
	&[\mathsf{B}]_{34} = - [\mathsf{B}]_{43} = \varOmega_- -\varOmega_+.
	\label{eq:redfield_matrix}
	\end{align}
\end{subequations}
All the remaining coefficients vanish.

The non-zero elements of the steady-state covariance matrix $ \mathsf{\Gamma}^\text{R} $ \textit{in the basis of the normal modes} $ \{ \pmb{\eta}_-, \pmb{\Pi}_-, \pmb{\eta}_+, \pmb{\Pi}_+ \} $ are
\begin{align}
\begin{split}
&[\mathsf{\Gamma}^\text{R}]_{11} = \frac{1}{\varOmega_-}\left(\frac{1}{2}+\langle \pmb{n}_- \rangle\right), \qquad [\mathsf{\Gamma}^\text{R}]_{22} = \varOmega_-\left(\frac{1}{2}+\langle \pmb{n}_- \rangle\right), \\
&[\mathsf{\Gamma}^\text{R}]_{33} = \frac{1}{\varOmega_+}\left(\frac{1}{2}+\langle \pmb{n}_+ \rangle\right),\qquad
[\mathsf{\Gamma}^\text{R}]_{44} = \varOmega_+\left(\frac{1}{2}+\langle \pmb{n}_+ \rangle\right), \\
&[\mathsf{\Gamma}^\text{R}]_{13} = [\mathsf{\Gamma}^\text{R}]_{31} =\frac{1}{2\sqrt{\varOmega_+\varOmega_-}}\langle \pmb{s}_{+-}\rangle, \qquad
[\mathsf{\Gamma}^\text{R}]_{14} = [\mathsf{\Gamma}^\text{R}]_{41} = -\frac{1}{2}\sqrt{\frac{\varOmega_-}{\varOmega_+}}\langle \pmb{d}_{+-}\rangle, \\
&[\mathsf{\Gamma}^\text{R}]_{23} = [\mathsf{\Gamma}^\text{R}]_{32} = \frac{1}{2}\sqrt{\frac{\varOmega_+}{\varOmega_-}}\langle \pmb{d}_{+-}\rangle, \qquad
[\mathsf{\Gamma}^\text{R}]_{24} = [\mathsf{\Gamma}^\text{R}]_{42} = \frac{1}{2}\sqrt{\varOmega_+\varOmega_-}\langle \pmb{s}_{+-}\rangle.
\end{split}
\end{align}
\label{eq:redfield_covariances}
Just like in Eq.~\eqref{eq:cov_global}, this can be rotated into the original quadratures by applying the suitable rotation matrix as defined in Eqs.~\eqref{eq:coordinates+-} and \eqref{eq:theta}.

Finally, the steady state heat currents obtained from the stationary solution of Eq.~\eqref{eq:partial_redfield} can be cast as
\begin{multline}
\dot{\pazocal{Q}}_c^\text{R} =-\dot{\pazocal{Q}}_h^\text{R} = \varOmega_+ \left[ W^c_{\varOmega_+}\langle \pmb{n}_+\rangle - W^c_{-\varOmega_+}(1+\langle \pmb{n}_+\rangle)  \right] + \varOmega_- \left[ W^c_{\varOmega_-}\langle \pmb{n}_-\rangle - W^c_{-\varOmega_-}(1+ \langle \pmb{n}_-\rangle) \right] \\
+ \frac{1}{2}\sqrt{\varOmega_+\varOmega_-}\langle \pmb{s}_{+-} \rangle \left[ (W^c_{\varOmega_-} - W^c_{-\varOmega_-})\cot{\vartheta} + (W^c_{\varOmega_+} - W^c_{-\varOmega_+})\tan{\vartheta} \right].
\end{multline}

%\bibliographystyle{apsrevfixedwithtitles}
%\bibliography{biblio}

%merlin.mbs apsrev4-1.bst 2010-07-25 4.21a (PWD, AO, DPC) hacked
%Control: key (0)
%Control: author (72) initials jnrlst
%Control: editor formatted (1) identically to author
%Control: production of article title (1) required
%Control: page (0) single
%Control: year (1) truncated
%Control: production of eprint (0) enabled
%

\end{document}